\documentclass[esd, manuscript]{copernicus}
\usepackage{amssymb}
\usepackage{amsmath}

\usepackage{graphicx}
\graphicspath{{./img/}}
\usepackage[normalem]{ulem}
\usepackage{xcolor}
\usepackage{cancel}
\usepackage{hyperref}
\usepackage{float}

\newcommand{\BB}{\mathcal{B}}
\newcommand{\FF}{\mathcal{F}}
\newcommand{\RR}{\mathcal{R}}

\nolinenumbers
\begin{document}

\title{Modeling the response of the marine carbon cycle to extreme CO$_2$ injection events }

\Author[1]{Punit}{Gandhi}
\Author[2]{Rowan}{Lockwood}
\Author[3]{Corinne}{Myers}
\Author[4]{Parimita}{Roy}
\Author[5][ivan.sudakow@open.ac.uk]{Ivan}{Sudakow}
\Author[6]{James}{Witts}
\Author[7]{Hao Helen}{Zhang}

\affil[1]{Department of Mathematics and Statistics, Virginia Commonwealth University, Richmond, Virginia, USA} 

\affil[2]{Department of Geology, William and Mary, Williamsburg, Virginia, USA} 

\affil[3]{Department of Earth and Planetary Sciences, University of New Mexico, Albuquerque, New Mexico, USA} 

\affil[4]{Department of Mathematics, Thapar Institute of Engineering and Technology, Patiala, Punjab, India} 

 \affil[5]{School of Mathematics and Statistics, The Open University, Milton Keynes, UK} 
 
 \affil[6]{The Natural History Museum, London, UK} 
 
\affil[7]{Department of Mathematics, University of Arizona, Tucson, Arizona, USA} 




\runningtitle{Carbon cycle response to extreme injection events}

\runningauthor{Gandhi et al.}

\received{}
\pubdiscuss{} 
\revised{}
\accepted{}
\published{}


\firstpage{1}

\maketitle

\begin{abstract}
We explore how the response of a conceptual model of the marine carbon cycle depends on the way in which carbon is injected from the atmosphere.  We find that, for single-injection pulses, the threshold amount required for a large response of the excitable system depends on pulse duration but not on its specific form.  We do, however, see differences in the number of large transient responses in carbon and, correspondingly, the duration of the response for different pulse shapes.  These differences are magnified as the system is pushed towards increased excitability and can be understood in terms of the geometry of an increasingly winding heteroclinic orbit.   
Inspired by Large Igneous Provinces (LIPs), we also consider random sequences of injection pulses. We find a wide range of possible responses for a given overall amount of injected carbon and duration, depending on the mean characteristics of the individual pulses. We also identify a resonance-like "Goldilocks" zone, in which intermediate pulse durations or arrival frequencies produce the largest
number of repeated transients, and we test the framework with illustrative scenarios motivated by the Siberian Traps and Columbia River Basalt Group.
\end{abstract}


\introduction  
\label{sec:intro}

Large igneous provinces (LIPs) represent periods of extreme volcanic activity resulting in 100,000s $km^3$ volume of igneous material emplaced over 100,000s $km^2$ area~\citep{BlackKarlstromMather2021,BRYAN2008175}. 
In addition to extruded lava, LIP volcanism also emits giga-tons of greenhouse gases and aerosols over 10s - 100s $kyrs$ per emplacement phase, resulting in the potential for significant climate impacts~\citep{BlackKarlstromMather2021,GrasbyBond2023,DeeganEtAl2023}. Given that LIP emplacement can occur over millions of years, these events and their Earth system consequences have been associated with profound changes in global biodiversity~\citep[e.g.,][]{GrasbyBond2023,Wignall2001,BondGrasby2017,ErnstYoubi2017,AlgeoShen2024,SUDAKOW202222}. Large igneous province volcanism is currently the consensus trigger for the end-Permian~\citep[e.g.,][]{VandeSchootbruggeWignall2016,BurgessMuirheadBowring2017,JoachimskiEtAl2020,KaihoEtAl2021,DalCorsoEtAl2022,WignallBond2024,BurgessBlack2025} and end-Triassic~\citep[e.g.,][]{VandeSchootbruggeWignall2016,MarzoliEtAl2017,SchoepferEtAl2022,KentEtAl2024,LiEtAl2026}
mass extinction events. It is also considered an exacerbating factor associated with the demise of the dinosaurs at the end-Cretaceous mass extinction, triggered by asteroid impact~\citep{courtillot2010cretaceous,keller2020mercury}, and more recently linked to mass extinction events at the end of the Ordovician~\citep[e.g.,][]{BondGrasby2017,JonesEtAl2017,QiuEtAl2025,YangEtAl2025}
and Late Devonian events~\citep{racki2020timing,racki2018mercury}. LIP volcanism is also heavily implicated as a causal mechanism for many of the ‘Oceanic Anoxic Events’ (OAE’s) in the Mesozoic~\citep{jenkyns2010geochemistry,walker2024oceanic}
and the Paleocene-Eocene Thermal Maximum during the early Cenozoic~\citep{kender2021paleocene,jones2023tracing}.

In these cases, a major pathway linking LIP activity to environmental stress is atmospheric CO$_2$ and other greenhouse-gas injection, followed by changes in ocean--atmosphere properties such as temperature, oxygenation, pH, and circulation. From a modelling perspective, LIP-driven CO$_2$ release is therefore naturally viewed as a time-dependent forcing of the marine carbon cycle, where the response depends not only on the total injected carbon but also on its rate, duration, pulse shape, and temporal clustering. This motivates the central question of this paper: how does the temporal structure of carbon injection determine the threshold, magnitude, and number of large excursions in an excitable marine carbon-cycle model?

We build on the conceptual marine carbon-cycle model proposed by~\citet{rothman2017thresholds,rothman2019characteristic}. On millennial to multimillion-year horizons, the upper ocean is treated as a weakly damped, open system involving alkalinity, dissolved inorganic carbon, and carbonate ion concentration, coupled by two opposing nonlinear feedbacks: carbonate burial and near-surface return. Weathering and degassing provide a nearly steady external supply, while air--sea exchange and interior mixing act as a damping channel. After reduction, the balances yield a smooth planar system with bounded, monotone process laws, forced by a time-dependent carbon injection rate. Independent evidence places the relevant carbon-cycle processes on distinct timescale bands, including carbonate compensation on millennial timescales, silicate-weathering stabilization on hundred-kiloyear timescales, and slower closure on multimillion-year scales~\citep{arnscheidt2022balance,arnscheidt2022presence,slyman2025tipping}. This reduced setting is simple enough to analyse geometrically, while retaining the key feature needed here: near an oscillatory transition, finite carbon perturbations can cross an excitability threshold and generate large transient excursions before the system relaxes back toward equilibrium.

Mathematically, the model has the phase-space structure of an excitable dynamical system. A physically interpretable steady state loses stability through an Andronov--Hopf bifurcation as burial and return sensitivities are varied; globally, a saddle-node of periodic orbits (SNPO) bounds a family of cycles. Between these loci lies a bistable wedge, where a stable fixed point and a stable limit cycle are separated by an unstable periodic orbit. Just outside this region, the system is \emph{excitable}: finite disturbances can cross a threshold and trace large, stereotyped transient excursions before returning to equilibrium, with amplitudes that are only weakly sensitive to the stimulus size once the threshold is exceeded and durations that are set by the nearby cycle period. This geometry is organized by a saddle point in the reduced phase portrait and an associated heteroclinic structure, which provide the phase-space mechanism for pulse-like responses and long relaxation tails~\citep{rothman2017thresholds,rothman2019characteristic}.

The same structure also connects the model to the standard distinction between bifurcation-, rate-, and noise-induced tipping. Slow drift of control parameters across the Hopf or SNPO boundaries can produce \emph{bifurcation tipping}, in which the attracting state changes between a fixed point and a periodic orbit~\citep{rothman2019characteristic}. Rapid or finite-duration carbon injection can instead produce threshold-crossing excitable responses; for ramps this may be interpreted as rate-dependent loss of tracking, while for pulses it reflects crossing of the excitable threshold~\citep{rothman2017thresholds,arnscheidt2022balance,wieczorek2011excitability,sudtip}. These deterministic responses, organized by the excitable geometry, are the main focus of the discussion below. Finally, in the bistable regime, adding stochastic variability would make the unstable periodic orbit a natural basin boundary for \emph{noise-induced tipping}, with escape governed by large-deviation and first-exit-time effects~\citep{rothman2019characteristic,slyman2025tipping}. We use this taxonomy to place the model in a broader tipping framework, while focusing in this paper on deterministic responses to ramped, pulsed, and randomly clustered CO$_2$ injection.

Building on this framework, our main contribution is to show how the temporal structure of carbon injection controls excitable carbon-cycle responses beyond the effect of total injected mass alone. We find that the threshold for producing a large excursion depends primarily on pulse duration and total injected carbon and is comparatively insensitive to whether the pulse is square or exponentially decaying. In contrast, the number of large excursions is highly sensitive to forcing duration and timing and is organized by the winding structure of the underlying heteroclinic orbit. This produces resonance-like tongues in the duration–mass plane and a non-monotone response to random pulse sequences, with the largest number of repeated excursions occurring for intermediate pulse sizes and arrival frequencies. These results provide a dynamical systems interpretation of how LIP-motivated styles of CO$_2$ release—including ramped, pulsed, and randomly clustered injection — can generate distinct marine carbon-cycle responses even for comparable total carbon input.

\section{A conceptual model for marine carbon cycle dynamics}

In this section, we establish our theoretical foundation by revisiting the marine carbon cycle model originally formulated by~\citet{rothman2017thresholds}, alongside a concise review of its fundamental properties as delineated in subsequent work~\citep{rothman2019characteristic}. While our analysis uses this exact mathematical framework, we explicitly distinguish our objectives from those in the prior literature. Rather than focusing solely on the total volume of injected mass, our primary contribution is to demonstrate that the specific temporal structure of carbon input dictates and controls the excitable responses of the marine carbon cycle. In doing so, we show that the rate and duration of injection play critical roles, independent of absolute mass thresholds.

On $10^3$--$10^6$\, yr horizons we treat the upper ocean as a weakly damped, open system with three mixed-layer reservoirs and two opposing feedbacks. Weathering/degassing supplies carbon and alkalinity at a nearly steady rate $J_{in}$ (optionally modified by a dimensionless imbalance $\nu$), while air-sea exchange and interior mixing relax dissolved inorganic carbon toward a background $W_0$ on a characteristic damping time $T_w$. Carbonate burial strengthens as the carbonate-ion level rises (stabilizing/compensating), whereas near-surface recycling back to DIC weakens as the carbonate-ion level rises. Here, we denote $A(t)$ as total alkalinity, $W(t)$ as dissolved inorganic carbon (DIC), and $C(t)$ as the carbonate-ion concentration that regulates $\mathrm{CaCO_3}$ preservation/burial. The system of differential equations describing their interaction is given by:

\begin{subequations}
\label{eq:model:roth:dim}
\begin{align}
\frac{d A}{dT} &= 
2 \bigg(
\underbrace{J_{in}}_{\text{alkalinity input}}
-
\underbrace{\BB(C)}_{\text{alkalinity loss (burial)}}
\bigg)
\label{eq:model:roth:dim:A}
\\[4pt]
\frac{d W}{dT} &= 
\underbrace{(1+\nu)\,J_{in}}_{\text{carbon input}}
-
\underbrace{\BB(C)}_{\text{burial sink}}
+
\underbrace{\RR(C)}_{\text{return to DIC}}
-
\underbrace{\frac{W-W_0}{T_w}}_{\text{relaxation}}
\label{eq:model:roth:dim:W}
\\[6pt]
\frac{dC}{dT} &\approx 
\underbrace{\FF(C)}_{\text{carbonate buffer}}
\left(\frac{dA}{dT}-\frac{dW}{dT}\right)
\label{eq:model:roth:dim:C}
\end{align}
\end{subequations}
where
\[
\BB(C)=B \frac{C^\gamma}{C^\gamma + C_p^\gamma},
\qquad
\RR(C)=R_\theta \frac{C_x^\gamma}{C^\gamma + C_x^\gamma},
\qquad
\FF(C)=F_0 \frac{C^\beta}{C^\beta + C_f^\beta}.
\]

The factor of 2 in $\dot A$ reflects the two alkalinity equivalents carried per mole of $\mathrm{CaCO_3}$. The buffer $\FF(C)>0$ represents fast carbonate acid--base equilibration and converts the instantaneous alkalinity--DIC imbalance into carbonate-ion tendency, $\dot C \approx \FF(C)(\dot A-\dot W)$; the imbalance $\nu$ perturbs only the carbon balance and will serve as our injection-rate control parameter below.

The process laws $\BB$, $\RR$, and $\FF$ are smooth, bounded, and monotone in $C$ ($\BB'(C)>0$, $\RR'(C)<0$). Together with linear relaxation in $\dot W$, these assumptions give a well-posed reduced system on the physical domain. The equivalent sign pattern for the reduced nondimensional Jacobian is given explicitly in Section~\ref{sec:excitability-model}, where it is used to locate the Hopf bifurcation.

Rescaling time by $T_w$ and measuring reservoir changes relative to fixed chemical crossovers yields, after eliminating $A$ via the buffer relation, a planar ODE in $(C,W)$ (smooth for $C>0$). In that nondimensional form, $\FF(C)$ acts as a positive gain, $\BB$ and $\RR$ appear as opposite-monotone, switch-like responses in $C$, and the $W$-equation retains unit linear damping. This scaling will be used in the subsequent stability and bifurcation analyses.

The amplitude $R_\theta=J_{in}\theta$ sets the strength of near-surface carbon return relative to external supply, with $C_x$ controlling where recycling turns off. Physically, a larger $\theta$ corresponds to a more ``leaky,'' recycling-dominated ocean state (e.g., warmer or more stratified surface waters, weaker particle ballasting), while a smaller $\theta$ corresponds to a more ``export-efficient'' state. Varying $\theta$ therefore represents physically plausible shifts in recycling strength and motivates the parameter studies below.

The presence of a single damping time $T_w$ implies a rate-mass dichotomy that structures later calculations: long forcings are limited by a critical \emph{rate}, while short pulses are limited by a critical \emph{mass} (expressed as a rate, this threshold decays approximately as $1/T$). This dichotomy is stated precisely, with explicit formulas, in Section~\ref{sec:excitability-model}.

\subsection{Reduced planar model}
Assuming equality in Eq.~\eqref{eq:model:roth:dim:C}, we can reduce to a planar system by solving in terms of $C$ and $W$. This closure is exact in the fast-buffer limit; the resulting $c=0$ point (see below) is an artifact of the reduced portrait and is not a true equilibrium of the full three-variable system (see Section~\ref{sec:excitability-model}). Converting to dimensionless variables $t$, $c$, and $w$ defined by
\[
T=T_w t,\quad C=C_pc,\quad W=W_0+J_{in} T_w w,
\]
the reduced planar system becomes
\begin{subequations}
\label{eq:model:roth}
\begin{align}
\frac{d c}{dt} &= \eta S_\beta(c,c_f) \bigg( 1 - \nu - b { S_\gamma(c,1)} - { \theta} { \overline{S}_\gamma(c,c_x)}  + w \bigg), \\  
\frac{d w}{dt} &= 1 + \nu - b {S_\gamma(c,1)} + { \theta} { \overline{S}_\gamma(c,c_x)}  - w.
\end{align} 
\end{subequations}
where 
\begin{equation*}
 { S_\alpha(c,c_0) } = \frac{c^\alpha}{c^\alpha + c_0^\alpha}, \qquad
{ \overline{S}_\alpha(c,c_0)} =1-S_\alpha(c,c_0)= \frac{c_0^\alpha}{c^\alpha + c_0^\alpha}.
\end{equation*} 
Here, we define dimensionless coefficients by $B=J_{in} b$, $R_\theta=J_{in}\theta$, $C_f=C_p c_f$, $C_x=C_p c_x$ and $\eta= J_{in}T_w F_0/C_p$; physically $F_0$ is the buffer strength and $\eta$ is the derived timescale ratio. We will take standard parameter values $b=4$, $\theta=5$, $c_f=0.399$, $c_x=0.5$, $\eta=1.577$, $\gamma=4$, $\beta=1.7$ and $\nu=0$ as a baseline for comparison. All of these numerical values---and the associated functional forms---are taken directly from the published model of~\citet{rothman2017thresholds} and are used here without modification. In the following sections, we numerically approximate solutions to the dimensionless system~\eqref{eq:model:roth} using the fourth-order, variable-step method (RK45) implemented in MATLAB's ODE suite~\citep{shampine1997matlab}. We report the results in the associated dimensional quantities, using the conversions summarized Table~\ref{tab:model:par}.  

\begin{table}
	\renewcommand{\arraystretch}{1.55}
	\centering
	\begin{tabular}{|p{2cm}|p{2cm}|p{2.5cm}|p{3cm}|p{4.5cm}|} 
		\hline
		 dimensionless quantity & baseline dimensionless value(s)& associated dimensional quantity & baseline dimensional value & description \\
		\hline
		\hline   
    $t$ & & $T=T_w t$ & $T_w=10^4\,yrs$ & time\\
    \hline
     $c$ & & $C=C_pc$ & $C_p=1.78\times 10^3\,Gt$ & oceanic carbonate \\
     \hline
     $w$ & & $W=W_0+J_{in}T_w w$ & $W_0=3.24\times10^4\,Gt$, $J_{in}=0.405\, Gt/yr$ & Dissolved Inorganic Carbon (DIC) in ocean \\
     \hline
     \hline
     $\eta$ & 1.577 & $F_0=C_p\eta/J_{in}T_w $ & $F_0=0.694$ & timescale ratio ($F_0$: carbonate buffering constant) \\
     \hline
     $c_f$ & 0.399 & $C_f=C_pc_f$ & $711\,Gt$ & buffer activation level \\
     \hline
     $c_x$ & 0.5 & $C_x=C_pc_x$ & $891\,Gt$ &respiration activation level\\
     \hline
     $b$ & 4 & $B=J_{in}b$ & $1.62\, Gt/yr$ &maximum burial rate \\
     \hline
     $\theta$ & 5 & $R_\theta=J_{in}\theta$& $2.025\,Gt/yr$ &maximum respiration rate \\
     \hline
    $\beta$ & 1.7 & $-$ & &buffer activation exponent\\
     \hline
    $\gamma$ & 4 & $-$ & &burial activation exponent \\
     \hline
     $\nu$ & 0 & $-$ & &carbon injection rate from atmosphere relative to baseline \\
     \hline
\end{tabular}
\caption{Variables and parameters of the model. All values are taken from~\citet{rothman2017thresholds}.
\label{tab:model:par}
}
\end{table}

\subsection{Remark on $\eta$}
The parameter
 $\eta=J_{in} T_w F_0 / C_p$
sets the timescale of carbonate-ion adjustment relative to mixed-layer DIC relaxation. Small $\eta$ corresponds to relatively fast relaxation in $w$ and slower evolution in $c$, as expected for well-ventilated, weakly buffered conditions; large $\eta$ corresponds to relatively fast carbonate response and slower evolution in $w$, as expected for sluggishly ventilated, strongly buffered or low preservation-threshold conditions. These two asymptotic limits provide useful interpretations of the model geometry: for $\eta\ll1$, trajectories rapidly approach the $w$-nullcline and drift in $c$, whereas for $\eta\gg1$, trajectories rapidly approach an algebraic constraint and relax in $w$~\citep{rothman2017thresholds,rothman2019characteristic}. Independent carbon-cycle timescale estimates place carbonate compensation at $\sim 10$~kyr and silicate-weathering stabilization at $\sim 200$--$400$~kyr, with amplified variance on multi-Myr horizons~\citep{arnscheidt2022balance,arnscheidt2022presence}. These timescale separations motivate the reduced damping structure of the model and provide context for the bifurcation regimes explored below. In the simulations below, however, we do not use either singular fast--slow limit. Throughout this paper we fix $\eta=O(1)$, using the baseline value in Table~\ref{tab:model:par}, so that both reservoirs evolve on comparable nondimensional timescales. This intermediate regime is the one in which the excitable structure central to our pulse-response results is most relevant.

\section{Excitability of the model and its dynamical features}
\label{sec:excitability-model}

\begin{figure}
    \centering
    \includegraphics[width=\linewidth]{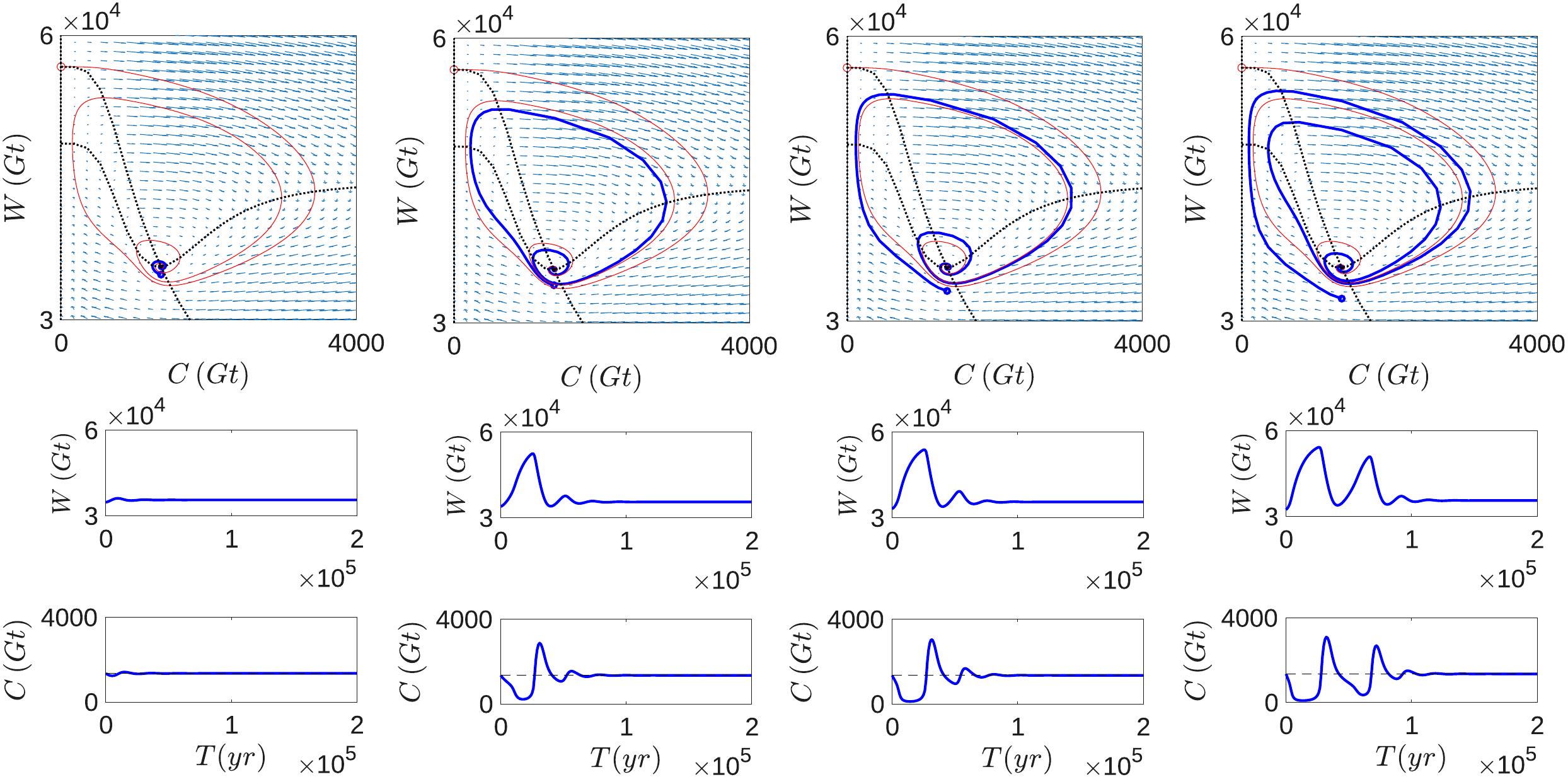}
    \caption{Phase portrait (upper) and time series (lower) for trajectories initialized at different distances below the stable equilibrium point at $(c,w)=(c^*,w^*-\Delta\nu)$ in excitable regime with $R_\theta=2.025\, Gt/yr \;(\theta=5)$. Nullclines are shown in dotted black and the vector field is indicated by thin blue arrows. In each case, the trajectory in phase space is shown in thick blue while the heteroclinic orbit connecting the two equilibria is shown in red.    We note shifting the $w$-component of the initial condition by $\Delta\nu$ is dynamically equivalent to increasing the parameter $\nu$ by $\Delta \nu$. 
    }
    \label{fig:model:phase}
\end{figure}

\begin{figure}
    \centering
    \includegraphics[width=\linewidth]{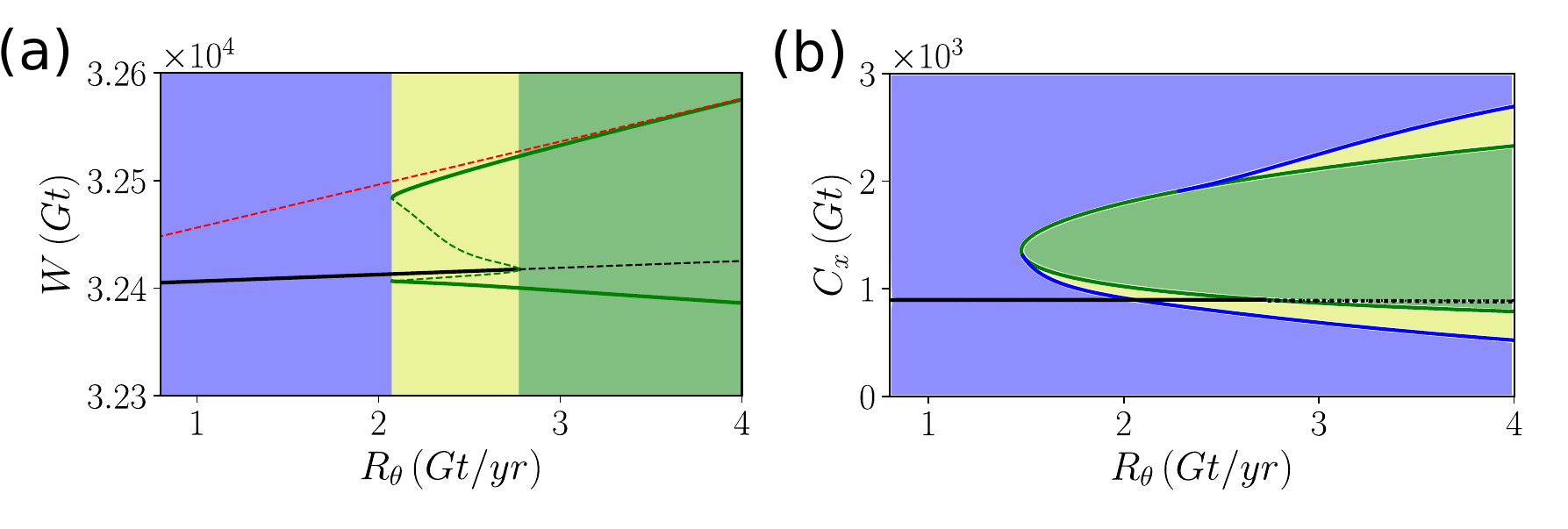}
    \caption{(a) Bifurcation diagram as a function of $R_\theta$ showing steady state equilibria in black and red, and min/max of periodic orbits in green. Stable and unstable states are indicated by solid and dashed lines respectively. (b) Two parameter bifurcation diagram in the $(R_\theta,C_x)$-plane indicating the Hopf curve in green and the saddle node of periodic orbits in blue. The steady state $(c^*,w^*)$ is unstable within the region enclosed by the Hopf curve on the right. Regions of bistability between periodic orbits and a steady state exist between the Hopf and SNPO curves. In the remainder of the parameter space, only a steady state is stable. The slice along $R_\theta$ with $C_x=891\, Gt$ ($c_x=0.5$) shown in (a) is indicated by the black line in (b). 
    }
    \label{fig:model:bif}
\end{figure}
\begin{figure}
    \centering
    \includegraphics[width=\linewidth]{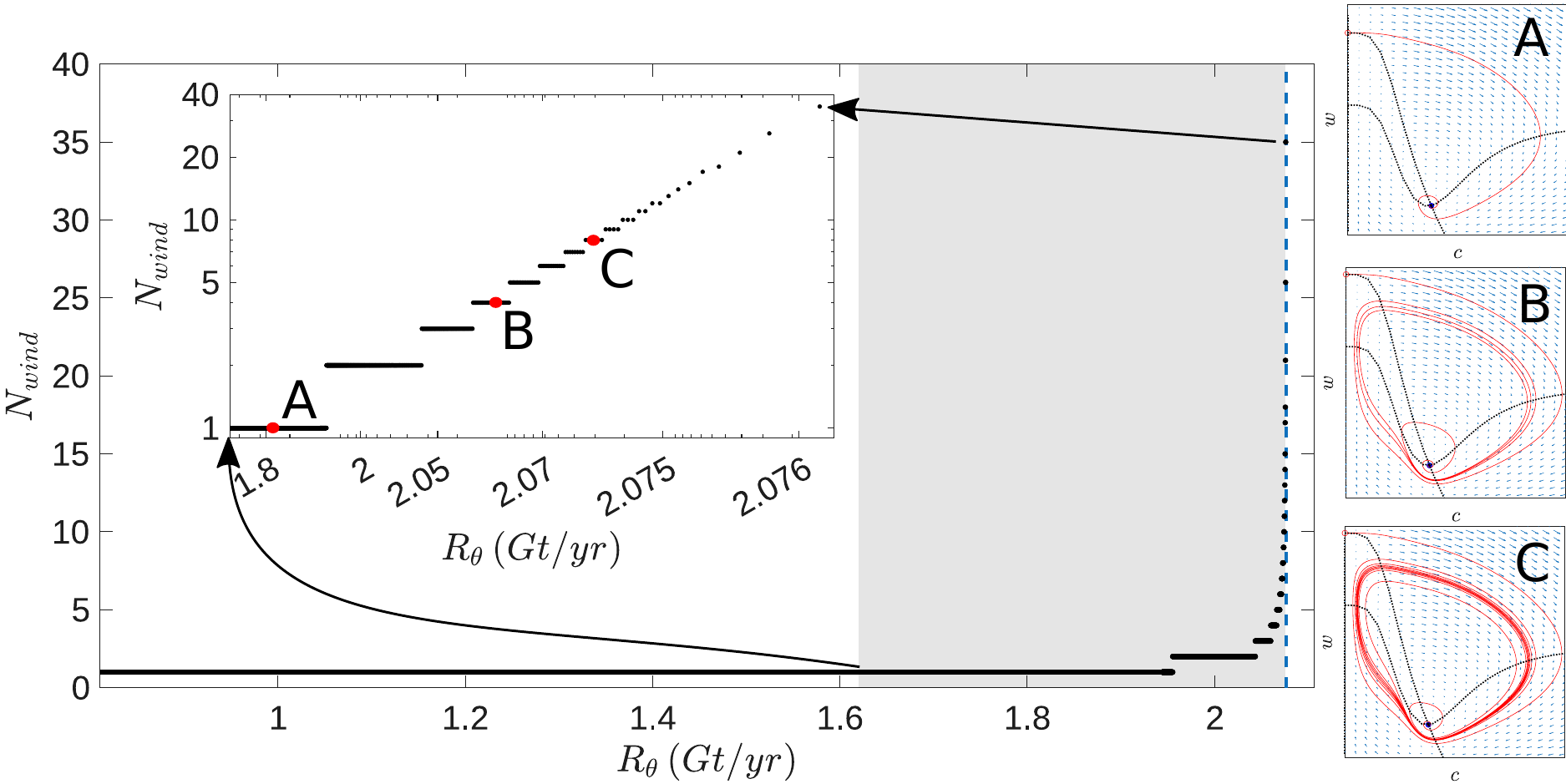}
    \caption{Number of large windings in heteroclinic orbit as a function of the parameter $R_\theta$.  Phase portraits illustrating the heteroclinic orbit are shown in the right panels for $R_\theta=1.823\, Gt/yr$ ($\theta=4.5$), $R_\theta=2.070\, Gt/yr$ ($\theta=5.11$), and $R_\theta=2.074\, Gt/yr$ ($\theta=5.12$). }
    \label{fig:model:Nwind}
\end{figure}

Building on the reduced planar system~\eqref{eq:model:roth}, we now characterize its equilibria, stability, and global bifurcation structure. The model is organized by a reduced saddle and an associated heteroclinic connection that generates large transient DIC anomalies after sufficiently strong perturbations. This structure provides the dynamical basis for the forcing experiments below and places the model within the standard taxonomy of bifurcation tipping, noise-induced tipping, and rate-induced tipping considered in Sections~\ref{sec:ramp}--\ref{sec:poissonpp}.

\subsection{Excitability} The reduced (dimensionless) system admits a physical steady state at $(c^*,w^*)$, where
\[
c^*=\frac{1}{(b-1)^{1/\gamma}},\qquad
w^*=\nu+\theta\,\overline{S}_\gamma(c^*,c_x)=\nu + \theta \frac{(b-1)c_x^\gamma}{(b-1)c_x^\gamma +1}.
\]
It also admits a second reduced fixed point with $c=0$ and $w^+=1+\nu+\theta$. In the full three-variable (dimensional) model this $c=0$ point is \emph{not} a true equilibrium because $dA/dT\neq 0$ when $C=0$, but it is dynamically useful in the reduced $(c,w)$ portrait: within the parameter range of interest it behaves as a saddle that organizes large transient responses \citep{rothman2019characteristic}.

Near oscillatory onset, the equilibrium $(c^*,w^*)$ lies in an \emph{excitable} regime: small perturbations decay, whereas finite kicks trigger large transient DIC anomalies associated with loops in phase space before the system returns to $(c^*,w^*)$ \citep{rothman2019characteristic}. The reduced saddle at $(0,w^+)$ together with a heteroclinic connection toward $(c^*,w^*)$ organizes both the number and persistence of these transient responses (Figure~\ref{fig:model:phase}). The reduced system is invariant under $(w,\nu)\mapsto(w+\Delta,\nu+\Delta)$; thus a step increase in $\nu$ is dynamically equivalent to a vertical shift of the initial condition---our standard pulse knob \citep{rothman2019characteristic}.
In this excitable regime, the peak DIC anomaly is largely \emph{insensitive} to how far the stimulus exceeds threshold, while the response duration is of the same order as the nearby limit-cycle period \citep{rothman2019characteristic}. This framing parallels classic \emph{excitable media}~\citep{cnsns2022excitable} and has been invoked for other Earth-system phenomena such as glacial cycles~\citep{rothman2019characteristic} and peatlands~\citep{wieczorek2011excitability}.

\subsection{Defining the transient-response count}
We quantify the response to a forcing episode by $N_{\mathrm{tr}}$, the number of large transient DIC anomalies in $w(t)$ before the system returns to a neighborhood of $(c^*,w^*)$. A large transient response is counted each time $w(t)$ exceeds the threshold
\[
w^*+\tfrac{1}{2}(w^+-w^*)
\]
following a return to within this band, where $w^+=1+\nu+\theta$ is the secondary (saddle) equilibrium. This threshold lies roughly midway between the stable equilibrium and the upper bound on $w$ set by the heteroclinic structure, and $N_{\mathrm{tr}}$ is insensitive to its precise value within a factor of $\sim\!2$. With the parameters given in Table~\ref{tab:model:par}, the threshold corresponds to a jump in DIC of $1.055\times 10^4$~Gt above the equilibrium value.

\subsection{Location of large transient responses in control space}
In the $(\theta,c_x)$ plane, the locus of large transient responses and the associated threshold $\nu_c$ are charted alongside the Hopf boundary, delineating an \emph{excitable wedge} adjacent to oscillatory onset \citep{rothman2019characteristic}.

\subsection{Bifurcations}
Linearization at $(c^*,w^*)$ yields a $2\times 2$ Jacobian $\mathcal{J}$ with
\[
\partial_w\dot w=-1,\qquad
\partial_w\dot c=\eta\,S_\beta(c^*,c_f)>0,\qquad
\partial_c\dot w=-\,b\,S'_\gamma(c^*,1)-\theta\,S'_\gamma(c^*,c_x)<0,
\]
and---using $\dot c=0$ at equilibrium---
\[
\partial_c\dot c=\eta\,S_\beta(c^*,c_f)\Big[-\,b\,S'_\gamma(c^*,1)+\theta\,S'_\gamma(c^*,c_x)\Big].
\]
Hence
\[
\mathrm{tr}\,\mathcal{J}=\eta S_\beta(c^*,c_f)\Big[-\,bS'_\gamma(c^*,1)+\theta S'_\gamma(c^*,c_x)\Big]-1,\qquad
\det \mathcal{J}=2\,\eta\,b\,S_\beta(c^*,c_f)\,S'_\gamma(c^*,1)>0.
\]
A Hopf bifurcation occurs when $\mathrm{tr}\,\mathcal{J}=0$ with $\det \mathcal{J}>0$; whether it is super- or subcritical depends on the first Lyapunov coefficient \citep{rothman2019characteristic}. Equivalently, as $c_x$ (or the effective return/preservation slopes) vary, the complex eigenvalues of $\mathcal{J}$ cross the imaginary axis---i.e., a Hopf onset consistent with Fig.~\ref{fig:model:bif} \citep{rothman2017thresholds}.

Figure~\ref{fig:model:bif}a shows a subcritical case at $c_x=0.5$. Figure~\ref{fig:model:bif}b maps the Hopf curve (green) and the saddle-node of periodic orbits (SNPO; blue) in $(\theta,c_x)$; between them lies a \emph{bistable wedge} (stable fixed point + stable cycle separated by an unstable cycle). Outside this wedge only a steady state is stable, and \emph{just outside} the SNPO boundary the system is \emph{excitable}: the vanished cycle leaves a pronounced ``ghost,'' producing large transient loops \citep{rothman2017thresholds,rothman2019characteristic}. We focus on the excitable band just below the SNPO at $\theta\approx 5.126$ (for the slice $c_x=0.5$ in panel a; see Figure~\ref{fig:model:Nwind} for the corresponding growth in the number of heteroclinic windings as this boundary is approached). Because the number of large transient DIC anomalies induced by a perturbation---the transient-response count $N_{\mathrm{tr}}$ 
---is highly sensitive near this boundary, subsequent figures sample several distinct values just below it ($\theta=4.0,\,4.5,\,5.0,\,5.1,\,5.11,\,5.12$) to illustrate how small changes in excitability alter the response.

\subsection{Bifurcation tipping}
Slow drift of $\theta$ or $c_x$ across the Hopf or SNPO curves in Fig.~\ref{fig:model:bif}b changes the attracting set between a fixed point and a periodic orbit~\citep{rothman2019characteristic}. We do not drive such slow parameter drift here; instead, the following sections hold $\theta$ (and hence the bifurcation structure) fixed and vary only the time-dependent forcing $\nu(t)$.

\subsection{Noise-induced tipping}
In the bistable wedge of Fig.~\ref{fig:model:bif}b, the basin boundary separating the fixed-point and limit-cycle attractors is the unstable periodic orbit $\Gamma$. Adding stochastic forcing to the reduced dynamics converts this into a noise-induced escape problem, with mean first-passage time governed by a Kramers-type law depending on the quasipotential barrier between the stable fixed point and $\Gamma$~\citep{slyman2025tipping}. This stochastic tipping regime is analyzed in detail by~\citet{slyman2025tipping} for the same underlying model; we do not pursue it further here, focusing instead on deterministic responses to different types of forcing.

\subsection{Rate-induced tipping and short-time law}
Even without crossing a static bifurcation, a sufficiently rapid driver can outpace tracking. Phanerozoic $\delta^{13}$C analysis yields a \emph{critical rate} for long forcings and a \emph{critical size (mass)} for short pulses; expressed as a rate, the short-pulse threshold scales approximately $\sim 1/T$---the \emph{rate--mass dichotomy} \citep{rothman2017thresholds,arnscheidt2022balance}. 
For short events of duration $t_i<\tau_w$, a compact form is
\[
\nu'_c(t_i)\;\approx\;\nu_c\,\frac{\tau_w}{t_i},
\]
and the corresponding \emph{critical mass} can be expressed as
\[
m_c\;\propto\;\nu_c\,J_{in}\,\tau_w,
\]
with constants as in the original derivation by~\citet{rothman2019characteristic}.  This short-time law is tested directly in Section~\ref{sec:ramp} using a
linear ramp in $\nu$, and extended in Sections~\ref{sec:pulse}--\ref{sec:lips}
to square pulses, exponential pulses, and random pulse sequences not
previously considered in this framework.

\section{Rate-induced tipping: linearly increasing injection rate}
\label{sec:ramp}

We begin by considering a linearly increasing injection rate to a new constant value
$\nu=\Delta\nu_{\text{rmp}}$ over a time period of $\tau_{\text{rmp}}$, given by
\begin{align}  
    \nu_{\text{rmp}}(t)&=\begin{cases}
    0 &t < 0, \\
     \Delta\nu_{\text{rmp}} \dfrac{t}{\tau_{\text{rmp}}} &0 \le t<\tau_{\text{rmp}}, \\
    \Delta\nu_{\text{rmp}} &\tau_{\text{rmp}}  \le t .
    \end{cases}
    \label{eq:model:nu:rmp}
\end{align}
Mathematically, this provides an avenue for exploring rate-induced tipping in the system. In subsequent sections, we will consider injection events with instantaneous shifts in injection rate $\nu$, and the linear ramp provides some insights into the limitations of this assumption. In particular, we find that when the ramp occurs faster than the characteristic response time of the system, the tipping threshold aligns with that of an instantaneous jump.

\begin{figure}
    \centering
    \includegraphics[width=0.8\linewidth]{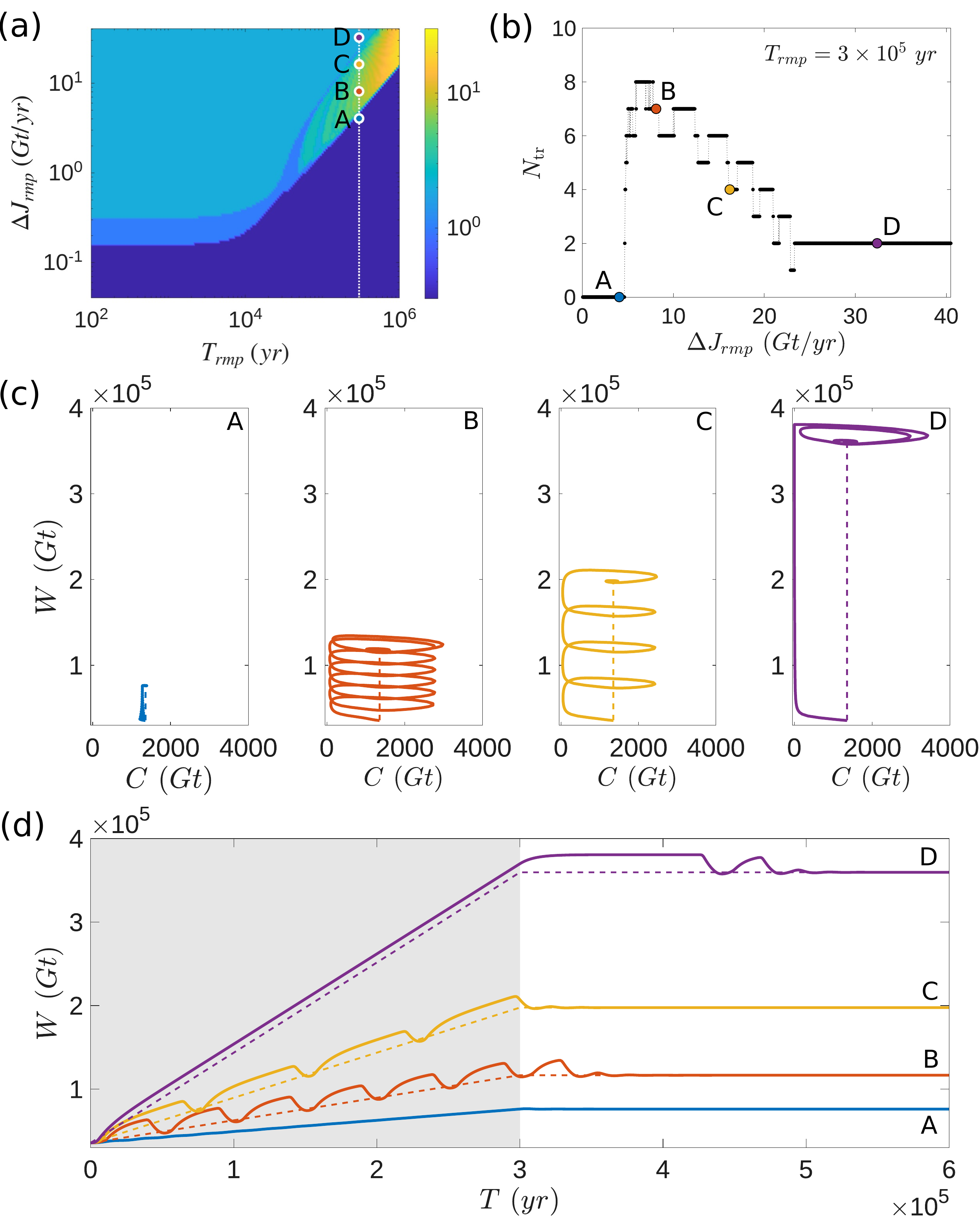}
    \caption{
    Linear ramp. (a) Transient-response count, $N_{\mathrm{tr}}$, during a linear ramp over a duration of $T_{\text{rmp}}$ and total change in carbon injection rate $\Delta J_{\text{rmp}}$. (b) Vertical slice from panel (a) along $T_{\text{rmp}}=3\times 10^5\,\mathrm{yr}$ showing $N_{\mathrm{tr}}$ as a function of $\Delta J_{\text{rmp}}$.
    (c) Phase portraits at locations along the slice indicated by colored dots to illustrate distinct dynamical behaviors. (d) Time series of DIC $W$ (solid lines) and equilibrium $W^*$ (dashed lines) corresponding to phase portraits above.} 
    \label{fig:rmp:tipping}
\end{figure}

Figure~\ref{fig:rmp:tipping} shows results from $10^6$-year ramped simulations initialized at the equilibrium point with $\nu=0$. The injection rate is linearly increased from $\nu=0$ to
$\nu=\Delta\nu_{\text{rmp}}$
over a period of $\tau_{\text{rmp}}$, and then remains constant for the remainder of the simulation. Panel (a) shows the
transient-response count, $N_{\mathrm{tr}}$,
observed during the simulation as a function of the ramp duration and the total shift in injection rate.

When the ramp occurs over a short enough period that the system does not have time to respond, it is like instantaneously moving the initial condition relative to the equilibrium point as was done in Figure~\ref{fig:model:phase}. In this case,
the tipping transition, as indicated by the threshold from $N_{\mathrm{tr}}=0$ to $N_{\mathrm{tr}}=1$, depends only on the change in injection rate $\Delta\nu_{\text{rmp}}$.
When the ramp occurs over a long enough period of time, the state is able to track the equilibrium point as it moves, and thus a larger shift in $\nu$ is required to cross that tipping threshold.

Figure~\ref{fig:rmp:tipping}(b) shows the slice along a constant ramp duration of $T_{\text{rmp}}=3\times10^5\,\mathrm{yr}$, as indicated by the vertical white line in panel (a).
Note that the $\Delta J_{\text{rmp}}$ axis is on a linear scale in panel (b), as opposed to the logarithmic scale used in panel (a).
Here we see an intricate structure in the transient-response count above the tipping threshold.
Interestingly, $N_{\mathrm{tr}}$ is maximized just above, but not far above, the tipping threshold.
Figure~\ref{fig:rmp:tipping}(c,d) show example phase-space trajectories and $w$ time series at different values of
$\Delta J_{\text{rmp}}$
along this slice in order to illustrate the range of observed dynamics.

\section{Excitable response to carbon injection pulses}
\label{sec:pulse}

In the following, we assume that the carbon injection rate is time-dependent, $\nu=\nu(t)$, as a way to model LIPs, impact-related perturbations, and other extreme carbon-injection events affecting the marine carbon cycle. In particular, we consider injection events that begin with a sufficiently rapid increase in $\nu$ that they can be treated as instantaneous on the response timescale of the reduced system. Large transient DIC anomalies of the type shown in Figure~\ref{fig:model:phase} can be induced by these injections, depending on their magnitude and duration. We are interested in how many such transient responses are induced by a given injection event.

We consider two distinct types of pulses:  a constant ``square'' injection pulse and an exponentially decaying injection pulse. In both cases, the pulse is initiated through an instantaneous increase in atmospheric carbon injection rate $\nu$, and we explore how the form of the pulse affects the dynamics for a given total carbon injection mass.

\begin{itemize}
\begin{subequations}
\label{eq:model:nu}
    \item[Case I:] Constant injection of $m$ total carbon over a duration of $\tau_{\text{sqp}}$
\begin{align}\label{eq:model:nu:sqp}
    \nu_{\text{sqp}}(t) &=\begin{cases}
    0 &\qquad  t < 0 \\
    \frac{m}{\tau_{\text{sqp}}} &0\le t<\tau_{\text{sqp}} \\
    0 &\tau_{\text{sqp}}  \le  t
    \end{cases}
    \end{align} 
\item[Case II:] Injection of $m$ total carbon with characteristic exponential decay time $\tau_{\text{exp}}$
\begin{align}\label{eq:model:nu:exp}
\nu_{\text{exp}}(t)&=\begin{cases}
    0   &t < 0 \\
    \frac{m}{\tau_\text{exp}} e^{-t/\tau_{\text{exp}}} &0  \le t 
    \end{cases}
    \end{align}
\end{subequations}
\end{itemize}

In Case I, the injection rate remains elevated at a constant value for a finite duration before instantaneously returning to its baseline value. Case II, on the other hand, allows the injection rate to decay back to baseline over a finite duration.

\subsection{Case I: Square injection pulse}
\label{sec:pulse:sqp}

The case of square injection pulses given in Eq.~\eqref{eq:model:nu:sqp} was studied by~\citet{rothman2019characteristic}, who provided a comprehensive theory for understanding the transition from small responses to large transient responses. If the injection occurs over a short enough time that the system cannot respond, the tipping threshold depends mainly on the total injected amount. If the duration of injection is long enough, the system has time to adjust and a larger injected amount is required to push it over the threshold. However, the transient-response count induced by the injection pulse has a more subtle dependence on pulse duration and magnitude.

\begin{figure}
    \centering
    \includegraphics[width=\linewidth]{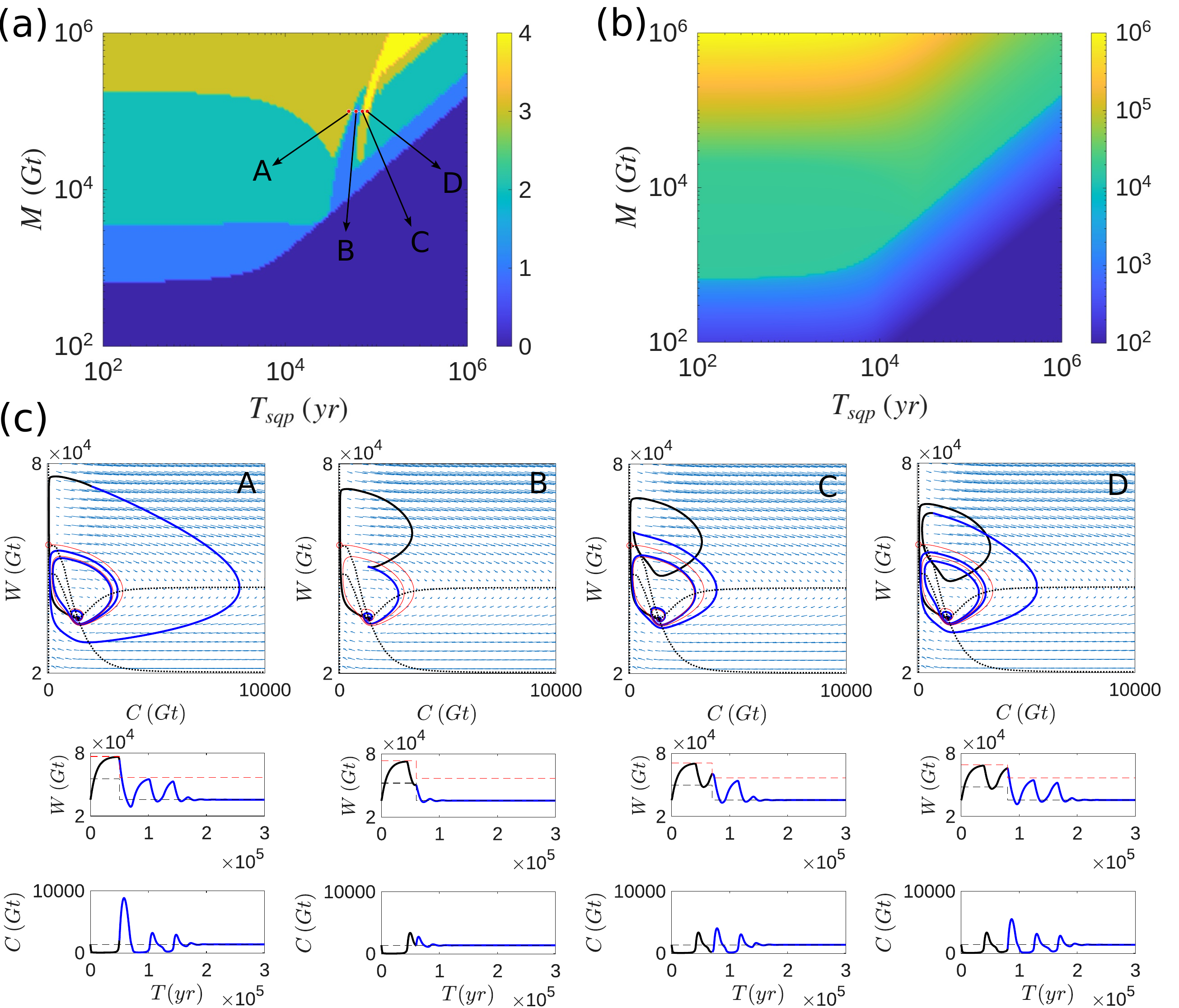}
    \caption{Square injection pulse. (a) Transient-response count, $N_{\mathrm{tr}}$, and (b) maximum jump in $W$, in units of Gt, from the equilibrium value as functions of pulse duration and total injection mass. (c) Phase portraits for parameter values along the white line in panel (a), showing the transition across a parameter region with non-monotone response structure. For the trajectories and time series, the injection pulse is on during the interval marked by the black solid line and off during the interval marked by the blue solid line. The heteroclinic orbit connecting the steady states is shown as a red solid line in the phase portraits. The red and black dashed lines in the time series indicate the unstable (red) and stable (black) quasi-static steady states for the fixed injection rate corresponding to the current value of $\nu$.}
    \label{fig:sqp:tipping}
\end{figure}

Figure~\ref{fig:sqp:tipping} summarizes results from simulations in which the system
is initialized
at the equilibrium with $\nu=0$ and undergoes a square injection pulse of total mass $m$ and duration
$\tau_{\text{sqp}}$.
As observed by~\citet{rothman2019characteristic}, the transition from
small responses to large transient responses, as indicated by the transition from $N_{\mathrm{tr}}=0$ to $N_{\mathrm{tr}}=1$
in Figure~\ref{fig:sqp:tipping}(a) follows the same qualitative pattern as observed for the linear ramp in Section~\ref{sec:ramp}. When the duration of the pulse is shorter than the characteristic response time of the system, this threshold
depends mainly
on the total mass $m$ injected. However, for
longer-duration pulses,
the system has time to adjust and a larger mass is required to
trigger a large transient response.
Figure~\ref{fig:sqp:tipping}(b) shows the
response magnitude, measured by the maximum shift in inorganic carbon $w$.
There is a
broad parameter range over which the peak response remains relatively constant, but significant jumps occur at larger injection masses.

While the behavior of the tipping transition is known~\citep{rothman2019characteristic,arnscheidt2022balance}, the severity of the response, measured by the transient-response count $N_{\mathrm{tr}}$, has a more subtle dependence on the parameters of the injection pulse. Figure~\ref{fig:sqp:tipping}(c) shows phase portraits of trajectories associated with simulations at fixed injection mass across a narrow range of pulse durations. Anywhere between $N_{\mathrm{tr}}=1$ and $N_{\mathrm{tr}}=4$ large transient responses are observed, and they do not vary monotonically with $\tau_{\text{sqp}}$. When $\tau_{\text{sqp}}$ is tuned so that the system has just completed one large transient response, the system is near a minimum at the time the pulse turns off and can return to equilibrium without executing additional large responses. On the other hand, when $\tau_{\text{sqp}}$ is tuned so that the system is near the peak of a large transient response when the pulse is turned off, up to two additional large responses can occur before the system returns to equilibrium. This resonance-like behavior is limited by the number of windings of the heteroclinic connection shown in red.

\subsection{Case II: Exponentially decaying injection pulse}
\label{sec:pulse:exp}

The exponentially decaying pulse given in Eq.~\eqref{eq:model:nu:exp} produces results qualitatively similar to those of the square pulse, indicating that the tipping threshold depends mainly on characteristic size and duration rather than on the detailed pulse shape. Quantitative differences remain, including a wider duration range over which only a single large transient response is observed.

\begin{figure}
    \centering
    \includegraphics[width=\linewidth]{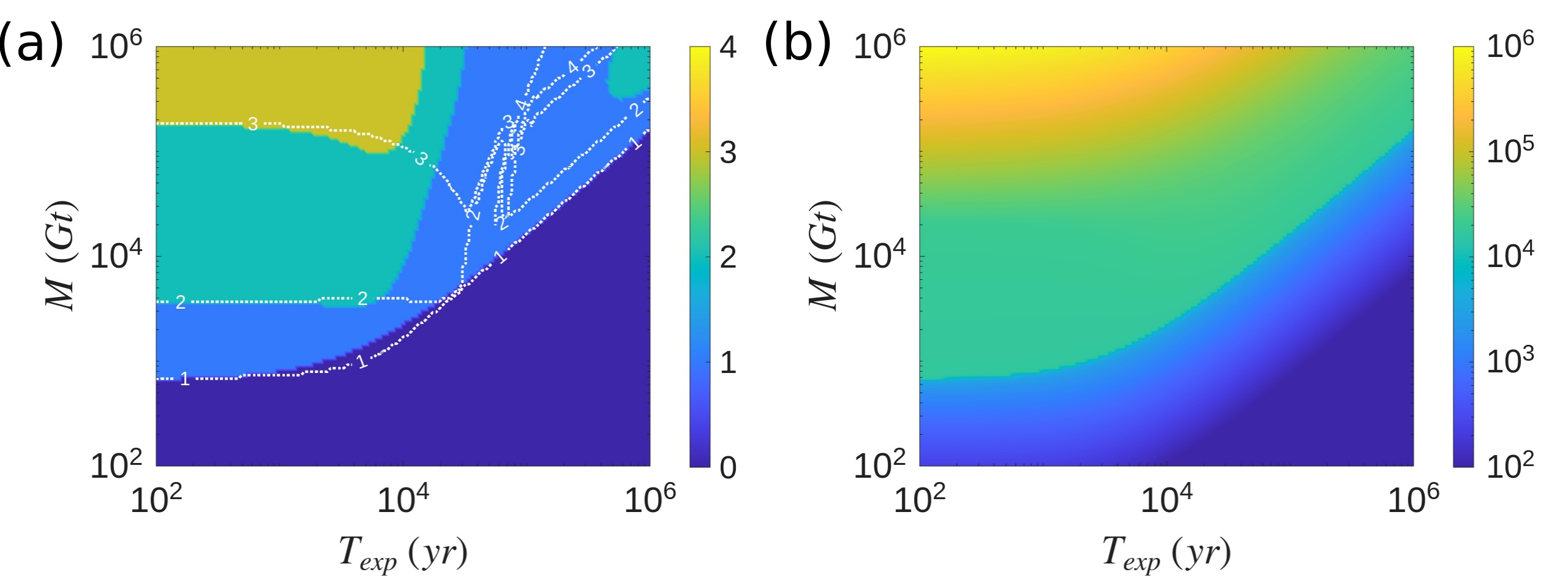}
    \caption{
    Exponentially decaying injection pulse. (a) Transient-response count, $N_{\mathrm{tr}}$, and (b) maximum shift in $W$, in units of Gt, relative to the equilibrium value as functions of pulse duration and total injection mass.}
    \label{fig:exp:tipping}
\end{figure}

Figure~\ref{fig:exp:tipping} shows results for simulations initialized at the equilibrium
with $\nu=0$ that
undergo
an exponentially decaying pulse with characteristic decay time
$\tau_{\text{exp}}$
and total injection mass $m$.
The tipping threshold---i.e., the transition from $N_{\mathrm{tr}}=0$ to $N_{\mathrm{tr}}=1$---is nearly identical to the threshold for the square injection pulse, as shown by the white dashed overlay indicating the analogous transition from Figure~\ref{fig:sqp:tipping}(a).
The reduced system~\eqref{eq:model:roth} is invariant under a constant shift of $w$, and a change in injection rate $\nu$ corresponds to a linear upward shift in $w^*$ while leaving $c^*$ constant.
We therefore interpret an instantaneous change in $\nu$ as a relative downward shift of the initial condition with respect to the equilibrium. The excitability observed in Figure~\ref{fig:model:phase} provides the mathematical mechanism for the responses to the injection pulses considered here.

While the tipping threshold remains relatively unchanged, the shifts in transitions associated with larger transient-response counts $N_{\mathrm{tr}}$ are more pronounced. The parameter region in which a single large transient response ($N_{\mathrm{tr}}=1$) is observed is considerably broader for the exponentially decaying pulse.

\section{Dependence on excitability of the system}
\label{sec:excitability}

We can characterize the excitability of the system by the number $N_{wind}$  of windings of the heteroclinic orbit connecting the steady states. As the parameter $\theta$ increases, we see an accumulation of windings as we approach the
saddle-node
of periodic orbits (Figure~\ref{fig:model:Nwind}).
We therefore use the maximum respiration parameter $\theta$ as a proxy for exploring how the excitability of the system influences the tipping dynamics of the system.

\begin{figure}
    \centering
    \includegraphics[width=\linewidth]{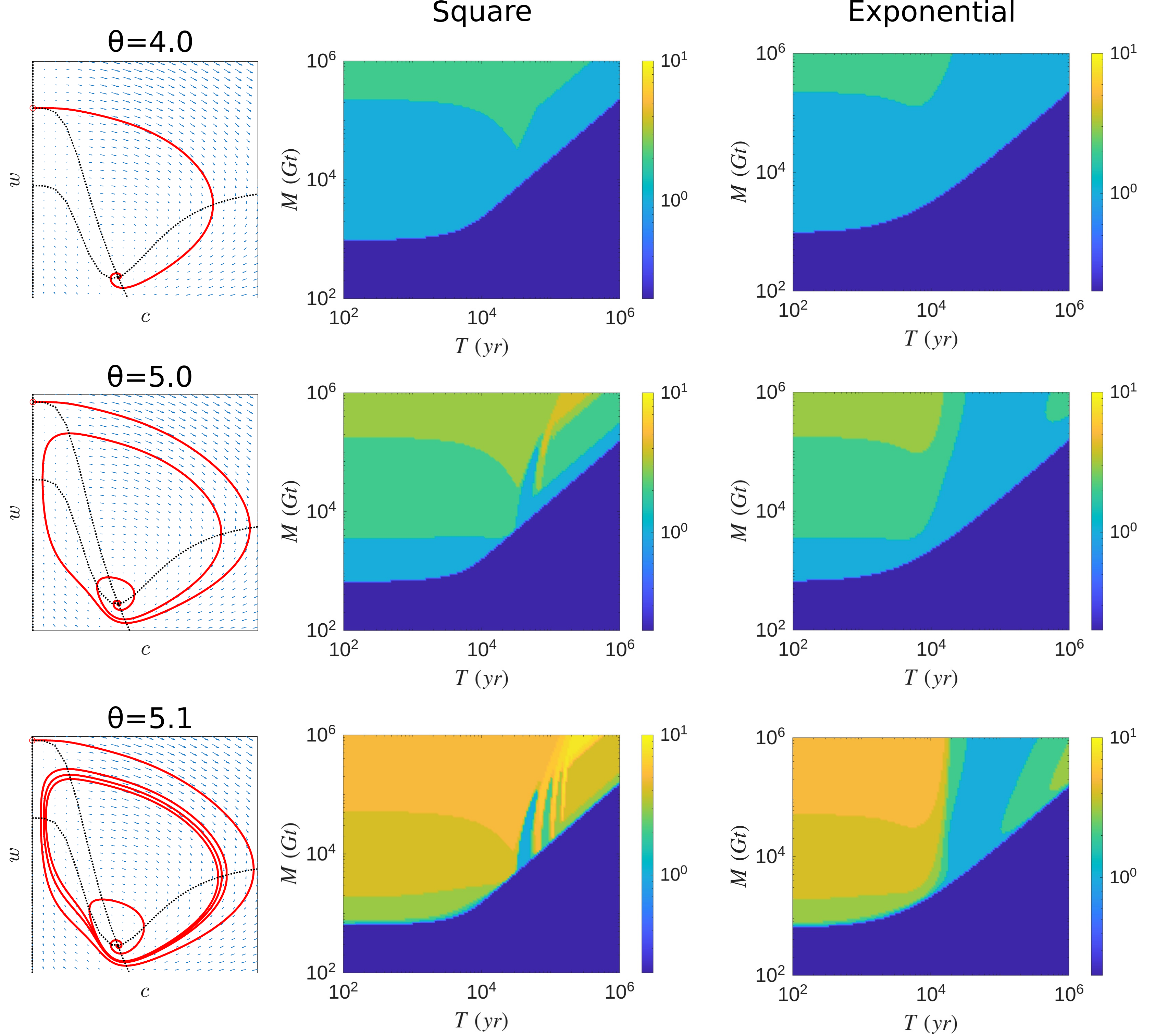}
    \caption{Dependence on excitability. Phase portraits illustrating the heteroclinic orbit are shown for $R_\theta=1.620\,\mathrm{Gt/yr}$ ($\theta=4.0$), $R_\theta=2.025\,\mathrm{Gt/yr}$ ($\theta=5.0$), and $R_\theta=2.066\,\mathrm{Gt/yr}$ ($\theta=5.1$). For each value of $\theta$, the remaining panels show the transient-response count, $N_{\mathrm{tr}}$, as a function of pulse duration $T$ and total injection amount for both square and exponentially decaying injection pulses.}
    \label{fig:th:tipping}
\end{figure}

Figure~\ref{fig:th:tipping} shows how changing $\theta$ changes both the heteroclinic geometry and the response to carbon-injection pulses. The phase portraits, shown from top to bottom for $\theta=4.0$ ($R_\theta=1.620$ Gt/yr), $\theta=5.0$ ($R_\theta=2.025$ Gt/yr), and $\theta=5.1$ ($R_\theta=2.066$ Gt/yr), correspond to increasingly excitable regimes. This is manifested in the increasing number of windings undertaken by the heteroclinic orbit shown in red. For each choice of $\theta$, the transient-response count, $N_{\mathrm{tr}}$, is shown as a function of injection mass and characteristic pulse duration for both square and exponentially decaying pulses. The center panels correspond to Figure~\ref{fig:sqp:tipping}(a) and Figure~\ref{fig:exp:tipping}(a) with $\theta=5.0$. When the system is less (more) excitable, we see a slight increase (decrease) in the injection mass required for the tipping threshold from $N_{\mathrm{tr}}=0$ to $N_{\mathrm{tr}}=1$. We also see an increase in the number of possible transient responses and more fine-scale structure as excitability increases. Both effects are direct consequences of the increased number of heteroclinic windings associated with greater excitability.

\section{Tipping statistics under random perturbations}
\label{sec:poissonpp}
We now explore how the characteristics of a random
sequence
of pulses
influence
the tipping dynamics for a given mean total injection rate over a fixed duration of time. Inspired by data about LIPs, we assume that, on average,
$10^5\,\mathrm{Gt}$
of carbon is injected through a sequence of random square pulses over a fixed duration of
$10^6\,\mathrm{yr}$.
We draw the duration
$\tau_{\text{sqp}}$
and injection mass $m$ for each pulse from exponential distributions with means
$T_r$ and $\bar m$.
We model the arrival times of these pulses by a Poisson point process and adjust the mean frequency of pulses so as to achieve the desired total mean injection rate for a given mean pulse injection mass
$\bar m$.

Specifically, for target total injection $M_{tot}$ over duration $T_{tot}$
and mean per-pulse mass $\bar m$, the arrival rate is set to
\[
\lambda = \frac{M_{tot}}{T_{tot}\,\bar m},
\]
so that the expected number of pulses is $\lambda T_{tot} = M_{tot}/\bar m$.

\begin{figure}
    \centering
    \includegraphics[width=\linewidth]{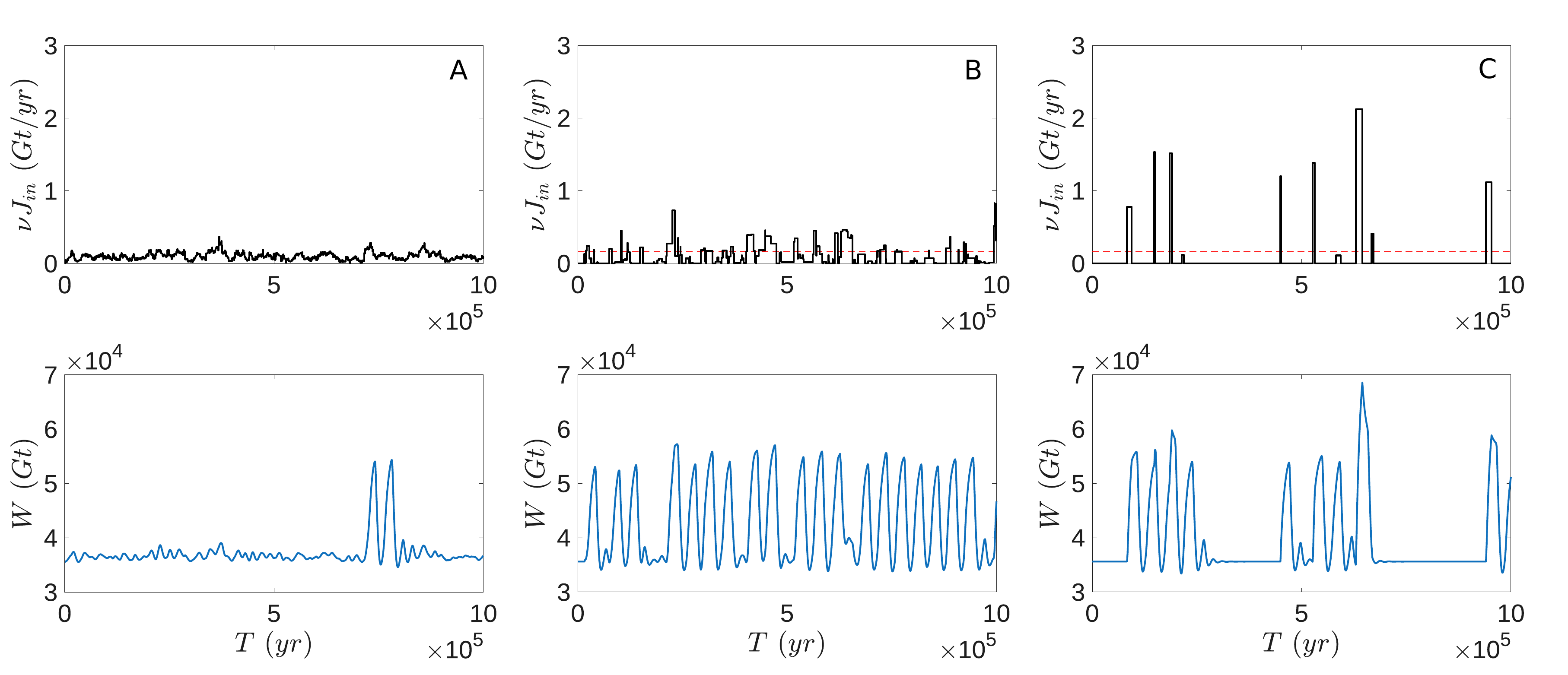}
    \caption{
    Time series of (top) $\nu J_{in}$ and (bottom) $W(t)$ for sequences of random injection events modeled by square pulses, with a total of $10^5\,\mathrm{Gt}$ of carbon injected above the baseline rate over $10^6\,\mathrm{yr}$. In each column, the mean size of the injection pulse varies from (A) $M=10^2\,\mathrm{Gt}$ to (B) $M=10^3\,\mathrm{Gt}$ and (C) $M=10^4\,\mathrm{Gt}$. The corresponding parameter locations are labeled in Figure~\ref{fig:PP:sq:mtTmu}(b). In all cases, the mean pulse duration is $T_r=10^4\,\mathrm{yr}$ and the mean pulse frequency is adjusted to keep the total expected injected carbon fixed. In the left panel, none of the injection events would induce a large transient response on its own, but the cluster of events near $T=7\times10^5\,\mathrm{yr}$ acts together to induce large transient DIC anomalies in $W$. In the right panel, there are 9 large transient responses induced by the random sequence, fewer than the 19 induced by the more frequent, lower-magnitude pulses in the middle panel.}
    \label{fig:PP:sq:timeseries}
\end{figure}

We first illustrate this random-forcing setup using example time series before turning to systematic parameter scans. Figure~\ref{fig:PP:sq:timeseries} shows three examples of injection $\nu$ and total DIC $w$ from simulations with random pulses. From left to right, the mean injection mass per pulse is
$M=10^2\,\mathrm{Gt}$, $10^3\,\mathrm{Gt}$, and $10^4\,\mathrm{Gt}$.
With a large number of
small pulses on average (left), each individual pulse cannot induce
large transient DIC anomalies
on
its
own. However, they overlap and work synergistically to induce two
large transient responses
at around $8\times10^5$ years into the simulation. The
larger pulses on average (right) are each typically capable of inducing
large transient responses,
but occur infrequently. We observe the most
large transient responses
with intermediate average pulse size. Many of the individual pulses are above the threshold for inducing
large transient responses,
and they occur frequently enough to work synergistically as well. As seen in the middle panels, this can lead to a nearly continuous period of
large transient DIC anomalies.

\begin{figure}
    \centering
    \includegraphics[width=\linewidth]{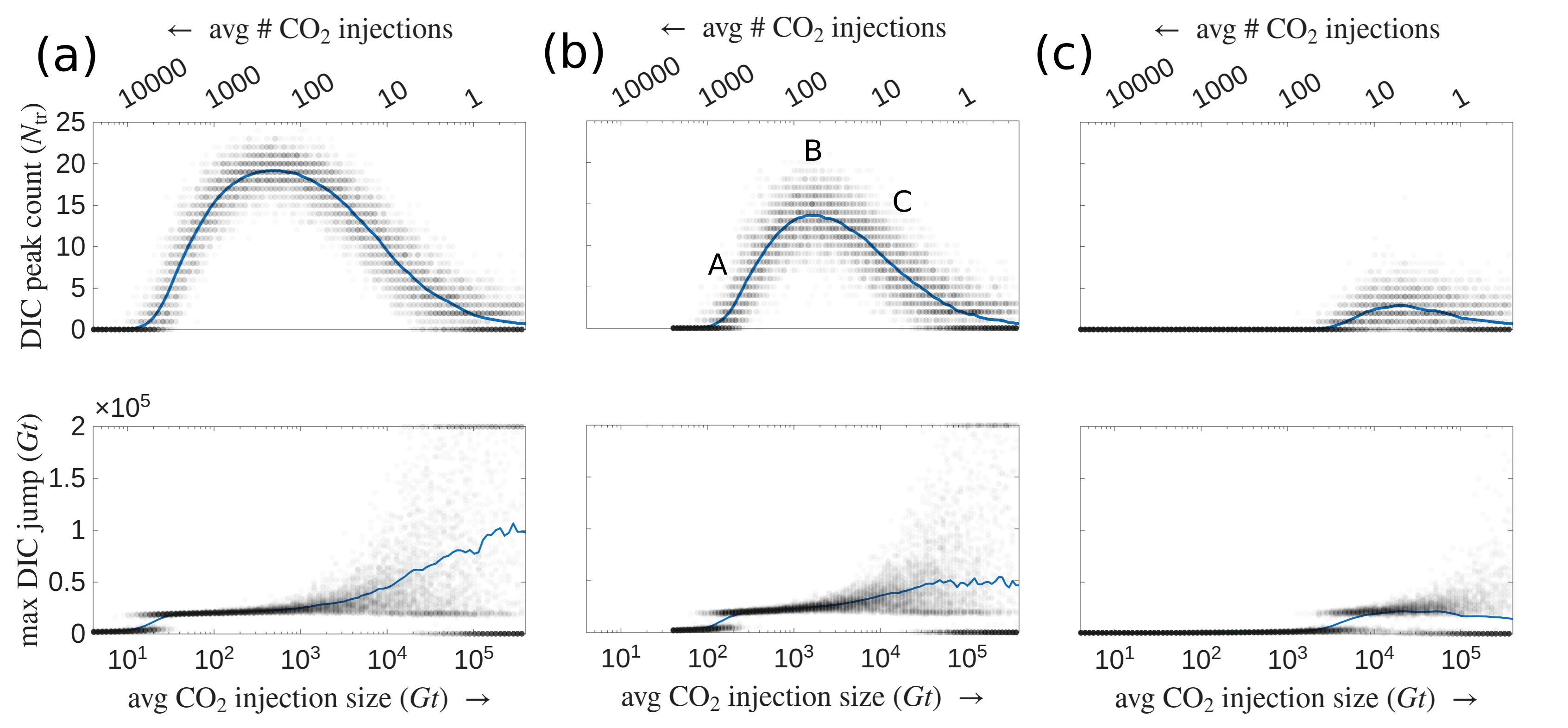}
    \caption{
    Transient-response count, $N_{\mathrm{tr}}$ (top), and magnitude of the largest DIC anomaly (bottom) as functions of the average size of the injection pulse for simulations in which a total of $10^5\,\mathrm{Gt}$ of carbon is injected above the baseline rate over $10^6\,\mathrm{yr}$. As the mean size of the injection pulse increases (lower horizontal axis label), the mean number of injection pulses decreases (upper horizontal axis label). The columns represent simulations with mean pulse duration (a) $T_r=10^3\,\mathrm{yr}$, (b) $T_r=10^4\,\mathrm{yr}$, and (c) $T_r=10^5\,\mathrm{yr}$. The results from 100 trials at each injection size are shown with gray dots and the averages are shown with blue lines. Any simulations with maximum DIC anomalies exceeding $1.055\times 10^4$~Gt are shown at the top of the graph in the lower panels. The points labeled in panel (b) correspond to parameter values for the time series from a single trial shown in Figure~\ref{fig:PP:sq:timeseries}.}
    \label{fig:PP:sq:mtTmu}
\end{figure}

The center panel of Figure~\ref{fig:PP:sq:mtTmu} shows results from simulations with random sequences of pulses in which, on average, $10^5\,\mathrm{Gt}$ of carbon is injected through random square pulses over a fixed duration of $10^6\,\mathrm{yr}$.
As with the trials shown in Figure~\ref{fig:PP:sq:timeseries}, the mean pulse duration of the cases shown in the center column
is
$T_r=10^4\,\mathrm{yr}$,
while the left and right panels consider pulses with mean duration of
$T_r=10^3\,\mathrm{yr}$ and $T_r=10^5\,\mathrm{yr}$,
respectively. For each case, 100 trials are simulated for each mean pulse injection mass and, as the mean mass increases, the mean number of injection pulses decreases as indicated by
the top and bottom
horizontal axes. The
transient-response count, $N_{\mathrm{tr}}$, in the top panels and the largest DIC anomaly from each trial in the bottom panels are shown
with gray dots, while a moving-window average over the trials is shown with a blue line.

The peak
transient-response count
occurs for intermediate pulse injection masses and therefore also intermediate pulse arrival frequencies across the different mean pulse-duration levels. The mean number of
large transient responses
decreases as the mean pulse duration increases, consistent with the observation that longer pulses require a higher threshold mass to induce
large transient responses.
While there may be a similar number of
large transient responses
induced by sequences of pulses with very large and very small mean injection size, the largest
DIC anomalies
are larger on average with very large injection masses.

\clearpage
\section{Application to LIPs}
\label{sec:lips}

In this section, we observe the fit between modeled CO$_2$  injection pulses and available empirical data from the geological record of LIPs. Specifically, we compare model predictions from the system~\ref{eq:model:roth} associated with the most biologically devastating LIP, the Siberian Traps~\citep[ST; emplaced at the Permo-Triassic boundary $\sim$252.2 – 251.4 Ma;][]{burgess2015high,BurgessMuirheadBowring2017}, 
and a much younger LIP, the Columbia River Flood Basalts~\citep[CRB; emplaced in the mid-Miocene $\sim$17.2-15.78 Ma;][]{kasbohm2023eruption,soderberg2024stratigraphy} a time interval in which no major biodiversity loss has been observed.
Both events led to changes in the global carbon cycle, including noted carbon isotope excursions. Whereas the ST LIP led to a mass extinction where 80--95\% of all life on Earth went extinct, no major extinction is associated with CRB emplacement. Here we use empirically derived records of both intrusive and extrusive LIP deposition, where volume and duration of igneous material produced are equated to relative CO$_2$ injection volume and duration; the ST are explored in Section~\ref{sec:lips:sib} and the CRB in Section~\ref{sec:lips:crb}.  

\subsection{Siberian Traps}
\label{sec:lips:sib}

To approximate carbon cycle systems behavior associated with ST emplacement, we model CO$_2$ injection as a random sequence of pulses as in Section~\ref{sec:poissonpp} that appear in three distinct phases as defined by previous researchers \citep{vasil2000evaluation,burgess2015high,BurgessMuirheadBowring2017}: (1) a pulsed injection phase lasting 330 kyrs, followed by (2) a steady injection phase lasting 420 kyrs, and (3) a final phase at the threshold between the pulsed and steady that lasts 150 kyrs.  We characterize pulsed vs. steady atmospheric CO$_2$ injection based on the ratio $F_{ps}=T_{sqp}/T_{\lambda}$ where $T_{sqp}$ is the mean duration of an injection pulse and $T_{\lambda}$ is the mean time between the pulse initiations.  We consider sequences with $F_{ps}<1$ as "pulsed" since, on average, there are quiescent periods without CO$_2$ injection between distinct injection pulses. Alternatively, sequences with $F_{ps}>1$ are considered "steady" since, on average, the injection pulses overlap.  Here, we take the mean pulse duration to be $T_{sqp}=20$ kyrs throughout and adjust $T_\lambda$ in each phase so that $F_{ps}=0.2$ in Phase 1, $F_{ps}=5$ in Phase 2, and $F_{ps}=1$ in Phase 3. We set the total mean injection to $M_{sib}=2\times10^5\,\mathrm{Gt}$ of additional CO$_2$ over the three phases.  We assume that Phase 1 injection $M_1$ is twice as much as Phase 3 injection $M_3$, based on the estimated volume of extrusive eruption \citep{vasil2000evaluation,burgess2015high,BurgessMuirheadBowring2017, BurgessBlack2025}, and therefore take $M_1=2M_3$.  Given that Phase 2 of ST emplacement is considered to be mostly intrusive \citep{BurgessMuirheadBowring2017, BurgessBlack2025}, there are no clear constraints on the relative amount of atmospheric injection associated with it.   We define the fraction of the Phase 1 and Phase 3 injection mass to the total injection mass as $f_p=(M_1+M_3)/M_{sib}$.

\begin{figure}
           \centering
        \includegraphics[width=\linewidth]{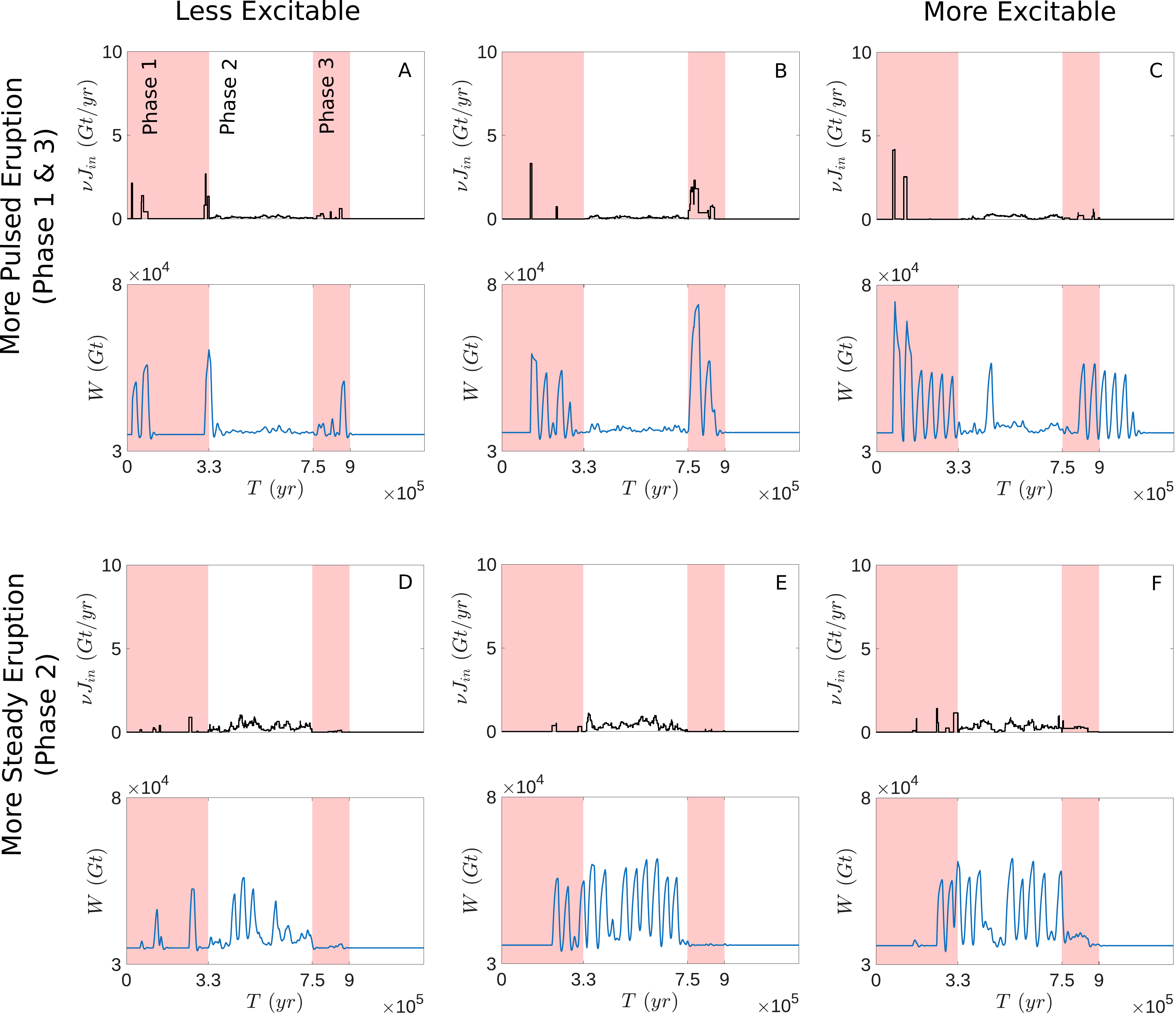} 
         \caption{Carbon injection scenarios inspired by the Siberian Traps.  Based on data and modeling from other studies, we take the total duration of the LIP at $9\times 10^{5}$ years and the mean total injection mass to be $2\times 10^5\; Gt$.  We assume three phases: the first lasts $3.3\times 10^5$ years with pulsed eruptions, the second lasts $4.2\times10^5$ years with steady eruptions, and the final phase lasts $1.5\times 10^5$ years with eruptions at a threshold between pulsed and steady.  We assume carbon is injected through a sequence of square pulses with arrival times modeled by a Poisson point process and exponentially-distributed amplitudes.  We consider different levels of excitability with, from left to right, $\theta=4,\,5,\, 5.1$. For each excitability level, we show the injection rate and total inorganic carbon $W$  from an example simulation that represents (top) pulsed eruptions and (bottom) steady eruptions.
         } \label{fig:lips:sib:ex}
        \end{figure}

\begin{figure}
           \centering
        \includegraphics[width=\linewidth]{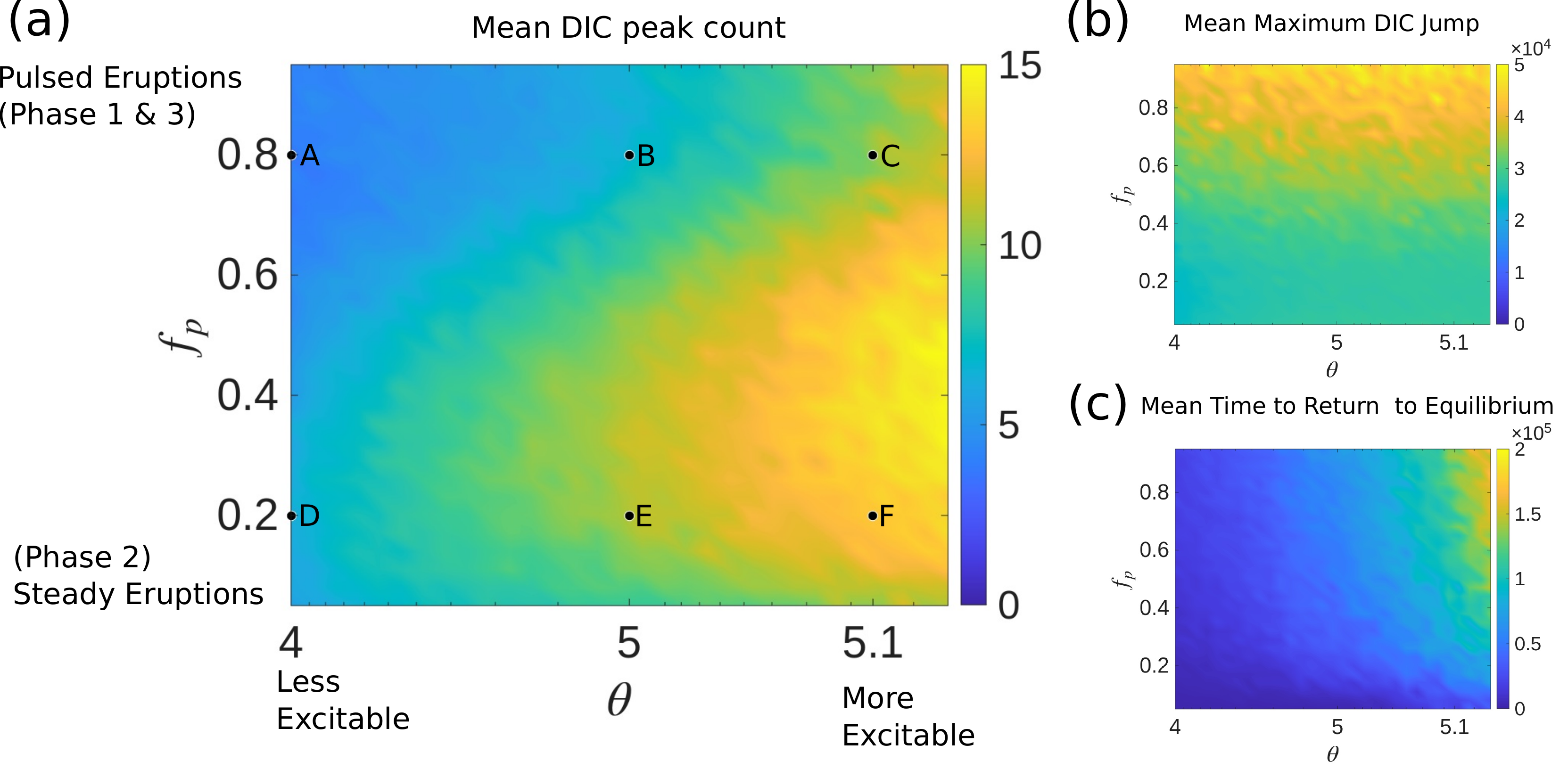} 
         \caption{Averages over 100 trials of Siberian Traps injection scenario on a 31$\times$ 31 grid in the $(f_p,\theta)$ parameter space.   Mean number of DIC peaks (left), mean maximum DIC jump (top right), and mean time to return to equilibrium after end of injection phases (bottom right) are shown.  The letters in the left panel correspond to parameters used in the example trials shown in Figure~\ref{fig:lips:sib:ex}. 
         } \label{fig:lips:sib:scan}
        \end{figure}

Figure~\ref{fig:lips:sib:ex} shows sample results for the random simulations inspired by Siberian Traps emplacement.  Each of the six pairs of panels shows time series of the injection rate $\nu J_{in}$ and the DIC $W$ for different CO$_{2}$ injection mass ratios $f_p$ and different excitability levels $\theta$.  In the top row (Figure~\ref{fig:lips:sib:ex}A-C), more CO$_2$ is injected through pulsed eruption ($f_p=0.8$), and in the bottom row (Figure~\ref{fig:lips:sib:ex}D-F), more CO$_2$ is injected in steady eruptions  ($f_p=0.2$). The level of excitability increases with $\theta=4.0$ (Figure~\ref{fig:lips:sib:ex}A,D), $\theta=5.0$ (Figure~\ref{fig:lips:sib:ex}B,E), and $\theta=5.1$ (Figure~\ref{fig:lips:sib:ex}C,F).    Figure~\ref{fig:lips:sib:scan}(a) shows averages over 100 trials of the total number of large DIC peaks produced during the simulation.  The locations in parameter space of the trials shown in Figure~\ref{fig:lips:sib:ex}A-F are labelled.  The maximum DIC peak size in $Gt$ and the time for the system to return to equilibrium after the end of Phase 3 are shown in Figures~\ref{fig:lips:sib:scan}(b) and~\ref{fig:lips:sib:scan}(c), respectively.    

Figure~\ref{fig:lips:sib:scan}(a) shows that when the injection mass is concentrated either in Phase 1 and 3 (pulsed CO$_2$ injection) or Phase 2 (steady injection), there is a decrease in the mean number of DIC peaks relative to a maximum number when emplacement style is a mix of both pulsed and steady CO$_2$ injection (e.g. $f_p\sim0.35$). Notably, if CO$_2$ injection is predominately pulsed or steady, the effect on the DIC peaks produced is not distinguishable between the two injection styles. However, consistent with the lower panels of Figure~\ref{fig:PP:sq:mtTmu}, the top right panel indicates that more pulsed eruptions lead to larger maximum values of DIC peaks on average compared to the steady injection scenario. Figure~\ref{fig:lips:sib:scan}(c) indicates that higher excitability ($\theta=5.1$) leads to a longer duration of carbon cycle system response before returning to equilibrium after CO$_2$ injection ceases on average.    
This is consistent with Figure~\ref{fig:lips:sib:ex}C,F which shows that an excitation in Phase 3 can lead to a larger number of DIC peaks and thereby, longer time for the system to return to equilibrium.   
        
\clearpage
\subsection{Columbia River Basalt Group}
\label{sec:lips:crb}

\begin{figure}
           \centering
        \includegraphics[width=\linewidth]{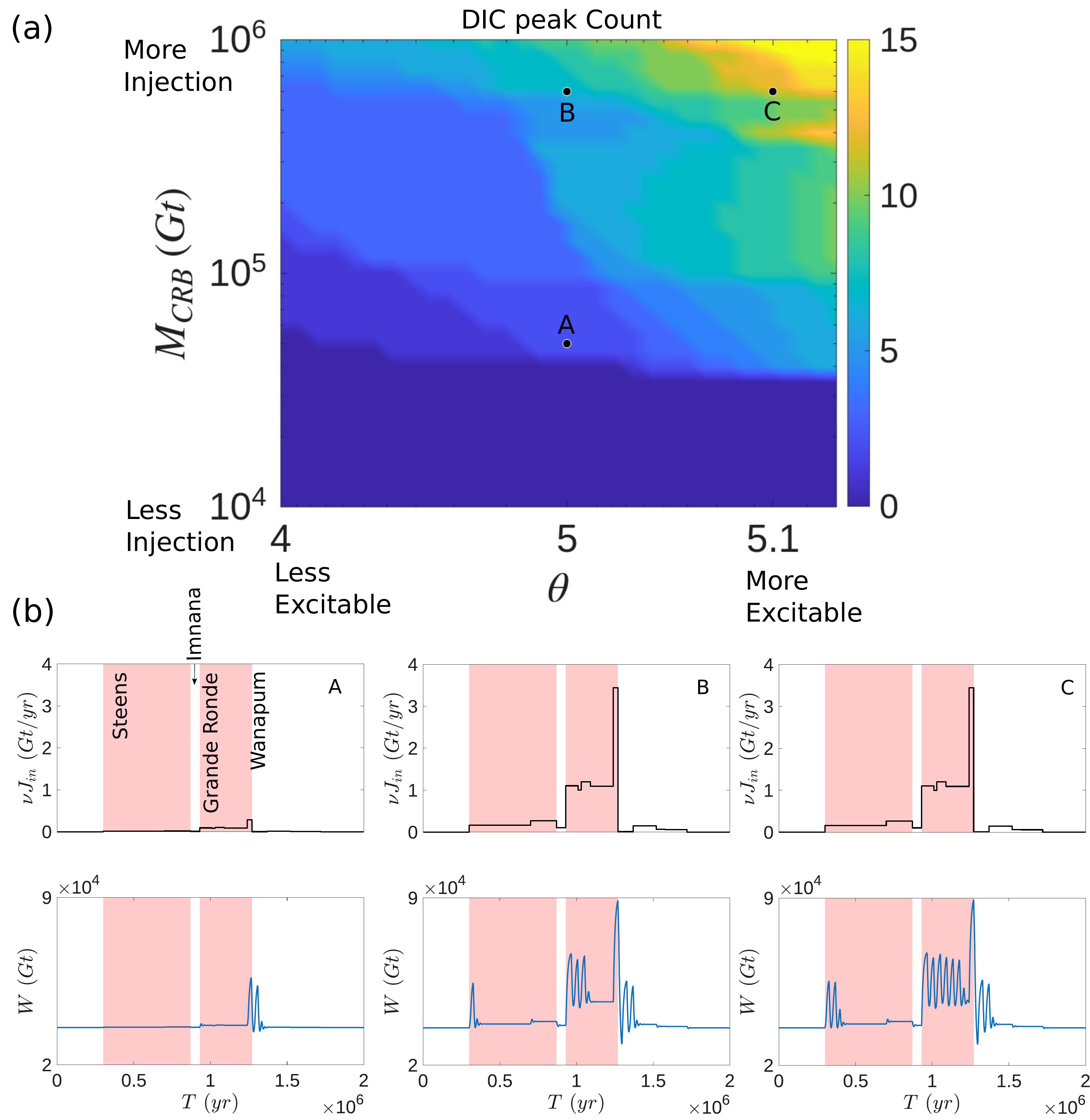} 
         \caption{Simulation results for deterministic sequence of square CO$_2$ injection pulses with timing, duration and relative injection amount based on Columbia River Basalt Group data shown in Table~\ref{tab:lips:crb}. (a) The transient-response count, $N_{\mathrm{tr}}$, induced by the sequence of pulses is shown as a function of the total injection amount $M_{CRB}$ and system excitability level $\theta$.  (b) Time series of the injection rate $\nu J_{in}$ and the DIC $W$ are shown for the corresponding parameter values labeled in (a).   } \label{fig:lips:crb}
        \end{figure}

To explore carbon cycle behavior associated with CRB emplacement, we make use of the timing, duration, and volume estimates of the main phase of Columbia River Basalt eruption lavas \citep[Table~\ref{tab:lips:crb};][]{kasbohm2023eruption,soderberg2024stratigraphy,kasbohm2018rapid}. Because eruption timing of these units is well-constrained, we use a deterministic sequence of square pulses, and assume that the CO$_2$ injection volume associated with each unit is proportional to the volume of extruded lava. Therefore, unlike with the Siberian Traps where both excitability ($\theta$) and the proportion of CO$_2$ injection was varied across the three emplacement phases with differing CO$_2$ injection styles (pulsed vs. steady), here we explore the effect of the total injection amount for a set of pulses of known proportion. Figure~\ref{fig:lips:crb}(a) shows the number of DIC peaks as a function of the total CO$_2$ injected  ($M_{CRB}$) and the excitability level $\theta$ of the marine carbon cycle model~\eqref{eq:model:roth}.  Figure~\ref{fig:lips:crb}(b) shows time series of the injection rate $\nu J_{in}$ and DIC $W$ for three parameter sets labeled in (a). Parameter set A=($M_{CRB}=5\times 10^4,\, \theta=5.0$) shows system behavior just above threshold for any excitations, and the high-peaked excitation pulse N2 at the end of the Grande Ronde period excites large DIC peaks.  For parameter set B=($M_{CRB}=6\times 10^5,\, \theta=5.0$), the total injection volume is increased with no change in $\theta$ and more of the injection pulses induce large DIC peaks.  
The effect of increased excitability is observed in parameter set C=($M_{CRB}=6\times 10^5,\, \theta=5.1$) as compared to B, which has the same injection mass but lower $\theta$ value. The same CO$_2$ injection pulses induce DIC peaks for both parameter sets B and C, but the number of peaks associated with these injection pulses is now increased.  This leads to a longer duration of system disequilibrium, and potential for approaching the widened resonance region for increased $\theta$ observed  in Figure~\ref{fig:th:tipping}.   That said, the number of large DIC peaks induced by the sequence of pulses does show a complicated structure in the $(\theta, M_{CRB})$ parameter plane.  In particular, increasing $M_{CRB}$ at higher excitability (e.g., $\theta=5.1$) can cause a decrease in the number of DIC peaks induced in some cases.  This is also a likely result of the resonance behaviors observed for square pulses in Section~\ref{sec:pulse:sqp} and~\ref{sec:excitability}.

\begin{table}
	\renewcommand{\arraystretch}{1.55}
	\centering
	\begin{tabular}{|r|c|c|c|}
		\hline
		 Name & Age ($Ma$) & Duration ($10^4\, yrs$) & Volume ($km^3$) \\
		\hline
		\hline  
  \multicolumn{4}{|l|}{Wanapum Basalt} \\  \hline
    Priest Rapids & 15.92 & 14 & 2,800\\ \hline
     Shumaker Creek, Roza & 15.98 & 6 & 1,300\\ \hline   
      Frenchman Springs & 16.13 & 15 &7,600\\ \hline      
    Eckler Mountain & 16.23 & 10 & 335\\ \hline\hline
    
     \multicolumn{4}{|l|}{Grande Ronde Basalt} \\  \hline
    N2 MSU &16.26 & 3 & 35,300\\ \hline
    R2 MSU &16.41 & 15 & 56,000\\ \hline
      N1 MSU & 16.47 & 6 & 24,600\\ \hline
    R1 MSU & 16.57 & 10 & 34,200\\ 
    \hline\hline
     Imnaha Basalt &16.8 & 31 & 11,000\\ \hline\hline
    Steens Basalt & 17.2 & 57 & 31,800\\\hline

\end{tabular}

\caption{Columbia River Basalt Formations~\citep{soderberg2024stratigraphy,camp2017field}.
\label{tab:lips:crb}
}
\end{table}

\clearpage
\section{Discussion}
\label{sec:discussion}

In this study we explored how the marine carbon cycle responds to different styles of atmospheric CO$_2$ injection by expanding upon the conceptual ocean--atmosphere carbon model of~\citet{rothman2017thresholds,rothman2019characteristic}. Our goal was to understand how perturbations associated with large igneous province (LIP) emplacement may disrupt the global carbon cycle and thereby affect climate and ocean chemistry. We focused on two measures of response: the transient-response count, $N_{\mathrm{tr}}$, defined as the number of large transient DIC anomalies produced before the system returns near equilibrium as a proxy for the duration of disequilibrium, and the peak change in dissolved inorganic carbon, $\max\Delta W$ (or $\max\Delta w$ in nondimensional variables), as a measure of the severity of the DIC anomaly. A central result is that these two measures are not controlled by the same aspects of the forcing. The number of DIC peaks is most sensitive to timing, recurrence, and system excitability, whereas the value of the DIC anomaly is more directly affected by the injected mass of CO$_2$ and injection pulse clustering.

The model operates in an excitable regime, in which a stable equilibrium coexists with a threshold for large transient responses. After an above-threshold perturbation, such as a CO$_2$ injection event, the system traces one or more large loops in phase space before returning toward equilibrium. These loops are organized by the heteroclinic structure of the reduced system, which connects the reduced saddle at $(0,w^+)$ to the stable equilibrium $(c^\ast,w^\ast)$~\citep{rothman2019characteristic}. The number of windings of this heteroclinic structure increases as the model approaches the saddle-node of periodic orbits (SNPO), allowing for larger DIC peaks and longer recovery times for a comparable perturbation. In physical terms, increasing the maximum respiration parameter $\theta$ corresponds to a more recycling-dominated or "leaky" ocean state (wherein near-surface recycling returns carbon to DIC more strongly at higher $\theta$ relative to ocean carbon burial). Although we cannot directly map $\theta$ to global temperature, warmer or more stratified ocean states may plausibly increase near-surface CO$_2$ recycling and therefore move the system toward greater excitability.

For the square injection pulse--in which the injection rate rises abruptly, remains constant for a prescribed duration, and then returns abruptly to baseline--our results reproduce the basic threshold behavior found in earlier work~\citep{rothman2017thresholds}. When the pulse is short, approximately $\lesssim 10^4$ yr in Figure~\ref{fig:sqp:tipping}, the key control is the total volume of CO$_2$ injected rather than the precise duration of the pulse. In this regime, large DIC peaks occur once a critical injected mass is exceeded. For longer pulses (Figure~\ref{fig:sqp:tipping}), the system has time to partially adjust during the forcing interval, and a larger total injection is required to cross the excitation threshold. This is threshold crossing under finite-duration forcing, not rate-induced tipping in the strict sense, because the injection rate changes discontinuously at the start and end of the pulse. Consistent with excitable dynamics, the peak DIC anomaly remains approximately stable over a wide range of above-threshold injection masses. For very large pulses, however, the system can be pushed beyond this characteristic excursion range, after which DIC relaxes rapidly rather than sustaining the elevated value.

A notable new result is the appearance of resonance-like behavior for square-pulse durations slightly exceeding $\sim 10^4$ yr. In this ``Goldilocks zone,'' the forcing is neither too short to act only as an instantaneous displacement nor so long that the system simply adjusts quasi-statically during the pulse. Instead, pulse duration becomes comparable to the intrinsic excursion and recovery time of the system. We observe narrow bands in the injection-mass--duration plane where only a single large excursion occurs even for above-threshold injections. This happens because the timing of pulse termination matters. If the pulse ends when the trajectory is already close to returning to equilibrium, the system relaxes after only one or a small number of loops. If the pulse ends while the trajectory is still far from equilibrium, additional DIC peaks follow. As the system becomes more excitable, additional Goldilocks bands appear, each associated with an additional winding of the heteroclinic orbit. Thus, pulse timing, not injected mass alone, controls the duration of disequilibrium in this regime.

Under the linear ramp scenario, the injection rate increases steadily over the ramp duration before leveling off. Here the perturbation threshold depends on both the ramp duration, which sets the rate of increase, and the total change in injection rate. For short ramps, again of order $\lesssim 10^4$ yr, the response resembles that of an instantaneous jump: the system has little time to adjust, and the threshold is close to the square-pulse threshold. For longer ramps, the state can partially track the moving quasi-static equilibrium, so a larger total injection is required to overcome the threshold. This is the clearest case of rate-induced tipping in the model because the response depends on whether the forcing changes too quickly for the system to track the moving state~\citep{wieczorek2011excitability,arnscheidt2022balance}. Within this rate-induced regime, we again observe Goldilocks behavior: the largest number of DIC peaks occur just above the threshold rather than far above it. At a fixed ramp duration, a forcing just above threshold can produce more peaks than a larger forcing that carries the system more directly toward a new elevated equilibrium.

This distinction is important for interpreting severity. In the model, a larger total injection does not always imply a longer disequilibrium interval. The highest DIC anomaly and the number of peaks can respond differently to the same forcing. A scenario with a high peak DIC value may produce a strong but relatively short disruption, whereas a near-threshold scenario may keep the system away from equilibrium through repeated peaks. Which of these is more damaging for marine ecosystems is beyond the scope of this conceptual model, but the result shows that duration and amplitude should not be treated as interchangeable measures of environmental stress.

The exponentially decaying pulse provides a useful comparison with the square pulse. In this case, the injection rate rises abruptly and then decays back to baseline over a characteristic decay time. The results are broadly consistent with the square-pulse case: the threshold for the first large DIC peak and the highest peak value are similar when total injected mass and characteristic duration are comparable. This supports the conclusion that the excitation threshold depends mainly on a characteristic size and duration, rather than on the detailed shape of the pulse. The main difference is that the exponential pulse lacks an abrupt termination. Because the forcing decays smoothly, there is no sharply defined pulse-off time to lock to the phase of the excursion. As a result, the narrow Goldilocks resonance bands seen for square pulses near $\sim 10^4$ yr are weakened or absent. This indicates that the resonance-like behavior is a timing effect associated with abrupt pulse termination, not simply a generic consequence of injection mass and duration.

We then considered random sequences of CO$_2$ injection pulses, motivated by geological evidence that LIP emplacement occurs through multiple eruptive and intrusive episodes rather than as a single smooth event. We modeled these sequences using square pulses with random durations, amplitudes, and arrival times. We focused on square pulses because the single-pulse experiments show similar threshold behavior for square and exponentially decaying pulses over short timescales, and because atmospheric CO$_2$ perturbations are redistributed on centennial-to-millennial timescales that are short compared with the $10^4$--$10^6$ yr DIC dynamics considered here~\citep{archer2009}. We describe sequences as ``pulsed'' when pulses are on average separated in time, and ``steady'' when pulses overlap strongly and behave more like a continuous background input. In these random forcing experiments, we again find a Goldilocks zone, now at the transition between pulsed and steady eruptive behavior. When pulses are very widely separated, each above-threshold event produces an isolated DIC peak, and the system has time to relax before the next event. When pulses overlap too strongly, the forcing becomes effectively steady and individual peaks are less distinct. The largest number of DIC peaks occurs between these limits, where pulses arrive often enough to re-perturb the system before it has fully returned to equilibrium, but not so often that the forcing becomes smooth. In other words, as with the original simulations above varying injection style, the system is again repeatedly disturbed just as it begins to recover. This keeps the carbon cycle in a near-continuous state of large transient DIC anomalies and produces the maximum in $N_{\mathrm{tr}}$ at intermediate pulse sizes and arrival frequencies.

The random pulse results also show that the number of DIC peaks alone does not fully distinguish pulsed from steady eruptive styles. Both very pulsed and very steady regimes can produce fewer repeated peaks than the transitional Goldilocks regime. However, the peak DIC anomaly value does distinguish these cases more clearly. For the same mean total CO$_2$ input and duration, pulsed sequences tend to reach larger peak DIC anomalies than steady sequences because isolated peaks can reach their full amplitude, whereas overlapping DIC peaks in the steady regime interfere, saturate, or merge into a smoother response. Increasing the excitability of the system by increasing $\theta$ does not necessarily increase the peak DIC anomaly, which is consistent with the excitable nature of the response, but it does increase the number of heteroclinic windings and the time required to return to equilibrium. Recovery time may therefore be a more sensitive indicator of system excitability than peak DIC alone.

Taken together, these results suggest that LIP-driven carbon cycle disruption should be interpreted as a dynamical forcing problem, not only as a carbon-budget problem. Total carbon release determines whether forcing is capable of crossing the excitation threshold, but the temporal structure of release determines how the system evolves after the threshold is crossed. Short pulses test the critical-mass regime; ramps test rate-dependent tracking; abrupt pulse termination reveals phase-locking to the excursion period; and random pulse trains show how recurrence time can resonate with the intrinsic recovery time of the carbon cycle. In this sense, the geological novelty of the results is also the mathematical novelty: LIP severity depends not only on forcing magnitude, but on the interaction between forcing timescale and the internal excitable geometry of the marine carbon cycle.

When applying the square pulse model to more realistic scenarios akin to the emplacement of the Siberian Traps (ST) and Columbia River Basalt (CRB) LIPs, we can observe potential system responses that may have contributed to the ST being one of the most deadly and the CRB one of the least deadly LIP events for global biodiversity. In the case of the ST, the geological record supports three main phases of LIP emplacement: Phase 1 and 3 were primarily extrusive with large basaltic flows erupted in discrete pulses; whereas Phase 2 was likely primarily intrusive with more steady emplacement of plutonic igneous deposits \citep{BurgessMuirheadBowring2017, BurgessBlack2025}. Phase length was based on this previous geological work, and total CO$_2$ injection was estimated injection was estimated using a compilation of sources~\citep[e.g.,][]{wu2023volcanic}. Under these conditions, varying the proportion of injected CO$_2$ in pulsed (Phase 1,3) versus steady (Phase 2) scenarios again demonstrated a Goldilocks zone at about 35\% of the total injection modeled as pulsed vs. steady (Figures~\ref{fig:lips:sib:ex},\ref{fig:lips:sib:scan}), wherein DIC peaks reached a maximum number. As total CO$_2$ injection increased or decreased away from this zone, DIC peak number was not distinguishable between pulsed vs. steady scenarios, however, the highest DIC peak anomaly values are only reached with > 35-percent pulsed CO$_2$ injection (Figure~\ref{fig:lips:sib:scan}). Modifying the excitability of the system ($\theta$) increases the time to return to equilibrium and also expands the Goldilocks region in which the largest number of DIC peaks is observed. We interpret this to suggest that were the Earth system to be more excitable \citep[e.g., if the Greenhouse world of the late Permian had associated higher $\theta$;][]{WignallBond2024,judd2024485,scotese2021phanerozoic}, then our model demonstrates an excellent pathway towards high magnitude and extended carbon cycle perturbation. This is observed in the geologic record of carbon stable isotopes at this time – wherein carbon isotope excursions are both large and continue for upwards of 5 Myrs post-ST event \citep{WignallBond2024,DalCorsoEtAl2022,payne2004large,zhang2018multiple}.

In the CRB simulations, geological data provide a good estimate of both the proportion and duration of CO$_2$ injection pulses over the main phases of CRB volcanism \citep{kasbohm2018rapid,kasbohm2023eruption, soderberg2024stratigraphy}. Using these boundary conditions, we varied the total CO$_2$ injected (maintaining duration/proportionality) and found, unsurprisingly, that higher total injected CO$_2$ increases the number of DIC peaks (Figure~\ref{fig:lips:crb}, points A-to-B). These peaks are produced because more emplacement phases are excited above background into producing DIC anomalies. When we increase system excitability ($\theta$), DIC peaks occur within the same emplacement phases, but at higher numbers that last longer (Figure~\ref{fig:lips:crb}, points B-to-C). This behavior again, moves CRB emplacement towards the Goldilocks zone in which system perturbations begin to amplify each other and prevent return to equilibrium. In the context of the Earth system at the time of CRB emplacement, system excitability was likely lower due to cooler global temperatures in the Cenozoic Icehouse~\citep{scotese2021phanerozoic,judd2024485}; this, combined with lower overall CO$_2$ injection volume could explain why this LIP event did not produce system disequilibria sufficient to result in mass extinction of the global biota. The forcing magnitudes used in these simulations are also consistent with estimates from more detailed LIP models. For a CAMP-like sequence of magma intrusions into carbon-bearing sedimentary rocks, \citet{Stewart2026} estimate peak thermogenic carbon release rates of approximately $5$--$130~\mathrm{Gt\,C\,yr^{-1}}$ and cumulative emissions of up to $\sim 8\times10^{3}~\mathrm{Gt\,C}$. Their thermal--kinetic model resolves discrete intrusion events rather than the idealized forcing profiles considered here, but the resulting rates and total carbon masses are of the same order as those explored in our simulations.

Several caveats are important. The model is conceptual and deliberately low-dimensional. It represents the marine carbon cycle through a reduced planar system with smooth saturating feedbacks, a single damping timescale, and idealized forcing histories. It does not include spatial structure, explicit ocean circulation, nutrient limitation, oxygen dynamics, ecological feedbacks, or detailed carbonate compensation chemistry. We also treat LIP-related CO$_2$ release as a forcing of the ocean--atmosphere carbon system rather than attempting to reconstruct a specific province in full geochemical detail. The parameter $\theta$ should be interpreted as a control on the strength of near-surface recycling relative to external supply, not as a direct temperature variable. Finally, the stochastic tipping problem is not analyzed here: adding noise in the bistable regime would make the unstable periodic orbit a basin boundary for noise-induced escape, but that large-deviation problem is left for future work~\citep{slyman2025tipping}. Alongside the possibility for an important role played by reduced $\theta$ as the global climate cooled in the ~between the ST and CRB eruptions, it is also important to note that fundamental changes occurred in the global carbon cycle over this time, notably the evolution of carbonate-shelled microplankton in the oceans~\citep{zeebe2003simple,henehan2023continental}, which provided additional complimentary pathways to mitigate dramatic carbon cycle changes caused by LIP eruptions~\citep{henehan2023continental}. 

Overall, however, the model predicts that the most persistent carbon-cycle disruption occurs when CO$_2$ release is neither too brief, too gradual, nor too smoothly distributed. Instead, the strongest repeated DIC peaks occur when injection pulses recur on timescales comparable to the characteristic recovery time of the carbon cycle. This provides a mathematical framework for comparing LIP  eruptive styles and other CO$_2$ injection events at fixed total carbon input and for assessing how the timing, duration, and clustering of CO$_2$ release may shape marine carbon cycle responses.




\codedataavailability{
All data used in this study are either generated by the model simulations described in the Article or derived from publicly available datasets cited in the manuscript. The model equations, parameter values, forcing scenarios, and numerical procedures needed to reproduce the simulations are provided in the Article.    Geological input data used to construct the Siberian Traps and Columbia River Basalt Group forcing scenarios are taken from the cited literature and summarized in the relevant tables. Code used to for simulations in  figures~\ref{fig:model:phase},\ref{fig:sqp:tipping},\ref{fig:PP:sq:timeseries}, and~\ref{fig:PP:sq:mtTmu} can be found at~\url{https://github.com/PunitGandhi/MarineCarbonCycleModel}. Other simulation outputs, processed figure data, and analysis scripts are available from the corresponding author upon reasonable request.
} 












\authorcontribution{
PG performed the numerical simulations, analyzed the results, and wrote the original draft. RL, CM, and JW contributed to writing and the paleodata synthesis and interpretation. PR contributed to the numerical simulations and mathematical interpretation. IS contributed to the mathematical interpretation, writing, and project administration. HHZ contributed to the data analysis methodology and mathematical interpretation. All authors contributed to the interpretation of the results, revised the manuscript, and approved the final manuscript.
} 

\competinginterests{The authors declare no competing interests.} 


\begin{acknowledgements}
This research was supported by the Mathematics of Mass Extinction research network (Ma(th)ssX), funded by the Isaac Newton Institute for Mathematical Sciences and EPSRC (Ref: EP/V521929/1). Part of this research was performed while the authors were visiting the Institute for Mathematical and Statistical Innovation (IMSI), which is supported by the National Science Foundation (Grant No. DMS-2425650).  This research is based upon work supported by the National Science Foundation, Division of Environmental Biology, Biodiversity on a Changing Planet program under Award number 2225013 to RL.

\end{acknowledgements}

\bibliographystyle{copernicus}
\bibliography{references}

@article{arnscheidt2022presence,
  title={Presence or absence of stabilizing Earth system feedbacks on different time scales},
  author={Arnscheidt, C. W. and Rothman, D. H.},
  journal={Sci. Adv.},
  volume={8},
  number={46},
  pages={eadc9241},
  year={2022},
  publisher={American Association for the Advancement of Science}
}

@article{arnscheidt2022balance,
  title={The balance of nature: A global marine perspective},
  author={Arnscheidt, C. W. and Rothman, D. H.},
  journal={Annu. Rev. Mar. Sci.},
  volume={14},
  number={1},
  pages={49--73},
  year={2022},
  publisher={Annual Reviews}
}

@article{courtillot2010cretaceous,
  title={Cretaceous extinctions: the volcanic hypothesis},
  author={Courtillot, Vincent and Fluteau, Fr{\'e}d{\'e}ric},
  journal={Science},
  volume={328},
  number={5981},
  pages={973--974},
  year={2010},
  publisher={American Association for the Advancement of Science}
}

@techreport{camp2017field,
  title={Field-trip guide to the vents, dikes, stratigraphy, and structure of the {C}olumbia {R}iver {B}asalt {G}roup, eastern {O}regon and southeastern {W}ashington},
  author={Camp, Victor E and Reidel, Stephen P and Ross, Martin E and Brown, Richard J and Self, Stephen},
  year={2017},
  institution={US Geological Survey}
}

@article{henehan2023continental,
  title={Continental flood basalts do not drive later {P}hanerozoic extinctions},
  author={Henehan, Michael J and Witts, James D},
  journal={Proc. Natl. Acad. Sci.},
  volume={120},
  number={21},
  pages={e2303700120},
  year={2023},
  publisher={National Academy of Sciences}
}

@article{jenkyns2010geochemistry,
  title={Geochemistry of oceanic anoxic events},
  author={Jenkyns, H. C.},
  journal={Geochemistry, Geophysics, Geosystems},
  volume={11},
  number={3},
  year={2010},
  publisher={Wiley Online Library}
}

@article{jones2023tracing,
  title={Tracing {N}orth {A}tlantic volcanism and seaway connectivity across the paleocene--eocene thermal maximum {(PETM)}},
  author={Jones, M. T. and Stokke, E. W. and Rooney, A. D. and Frieling, J. and Pogge von Strandmann, P. A. E. and Wilson, D. J. and Svensen, H. H. and Planke, S. and Adatte, T. and Thibault, N. and others},
  journal={Climate of the Past},
  volume={19},
  number={8},
  pages={1623--1652},
  year={2023},
  publisher={Copernicus Publications G{\"o}ttingen, Germany}
}

@article{judd2024485,
  title={A 485-million-year history of {E}arth’s surface temperature},
  author={Judd, Emily J and Tierney, Jessica E and Lunt, Daniel J and Monta{\~n}ez, Isabel P and Huber, Brian T and Wing, Scott L and Valdes, Paul J},
  journal={Science},
  volume={385},
  number={6715},
  pages={eadk3705},
  year={2024},
  publisher={American Association for the Advancement of Science}
}

@article{kasbohm2023eruption,
  title={Eruption history of the {C}olumbia {R}iver {B}asalt {G}roup constrained by high-precision {U-Pb} and {40Ar/39Ar} geochronology},
  author={Kasbohm, J. and Schoene, B. and Mark, D. F. and Murray, J. and Reidel, S. and Szymanowski, D. and Barfod, D. and Barry, T.},
  journal={Earth Planet. Sci. Lett.},
  volume={617},
  pages={118269},
  year={2023},
  publisher={Elsevier}
}

@article{kasbohm2018rapid,
  title={Rapid eruption of the {C}olumbia {R}iver flood basalt and correlation with the mid-{M}iocene climate optimum},
  author={Kasbohm, J. and Schoene, B.},
  journal={Sci. Adv.},
  volume={4},
  number={9},
  pages={eaat8223},
  year={2018},
  publisher={American Association for the Advancement of Science}
}

@article{keller2020mercury,
  title={Mercury linked to {D}eccan {T}raps volcanism, climate change and the end-{C}retaceous mass extinction},
  author={Keller, Gerta and Mateo, Paula and Monkenbusch, Johannes and Thibault, Nicolas and Punekar, Jahnavi and Spangenberg, Jorge E and Abramovich, Sigal and Ashckenazi-Polivoda, Sarit and Schoene, Blair and Eddy, Michael P and others},
  journal={Glob. Planet. Change},
  volume={194},
  pages={103312},
  year={2020},
  publisher={Elsevier}
}

@article{kender2021paleocene,
  title={{P}aleocene/{E}ocene carbon feedbacks triggered by volcanic activity},
  author={Kender, S. and Bogus, K. and Pedersen, G. K. and Dybkj{\ae}r, K. and Mather, T. A. and Mariani, E. and Ridgwell, A. and Riding, J. B. and Wagner, T. and Hesselbo, S. P. and others},
  journal={Nat. Commun.},
  volume={12},
  number={1},
  pages={5186},
  year={2021},
  publisher={Nature Publishing Group UK London}
}

@article{payne2004large,
  title={Large perturbations of the carbon cycle during recovery from the end-{P}ermian extinction},
  author={Payne, Jonathan L and Lehrmann, Daniel J and Wei, Jiayong and Orchard, Michael J and Schrag, Daniel P and Knoll, Andrew H},
  journal={Science},
  volume={305},
  number={5683},
  pages={506--509},
  year={2004},
  publisher={American Association for the Advancement of Science}
}

@article{racki2020timing,
  title={Timing of dicynodont extinction in light of an unusual Late Triassic Polish fauna and Cuvier’s approach to extinction},
  author={Racki, G. and Lucas, S. G.},
  journal={Hist. Biol.},
  volume={32},
  number={4},
  pages={452--461},
  year={2020},
  publisher={Taylor \& Francis}
}

@article{racki2018mercury,
  title={Mercury enrichments and the Frasnian-Famennian biotic crisis: A volcanic trigger proved?},
  author={Racki, Grzegorz and Rakoci{\'n}ski, Micha{\l} and Marynowski, Leszek and Wignall, Paul B},
  journal={Geology},
  volume={46},
  number={6},
  pages={543--546},
  year={2018},
  publisher={Geological Society of America}
}

@article{rothman2019characteristic,
  title={Characteristic disruptions of an excitable carbon cycle},
  author={Rothman, D. H.},
  journal={Proc. Natl. Acad. Sci.},
  volume={116},
  number={30},
  pages={14813--14822},
  year={2019},
  publisher={National Acad Sciences}
}

@article{rothman2017thresholds,
  title={Thresholds of catastrophe in the {E}arth system},
  author={Rothman, D. H.},
  journal={Sci. Adv.},
  volume={3},
  number={9},
  pages={e1700906},
  year={2017},
  publisher={American Association for the Advancement of Science}
}

@article{shampine1997matlab,
  title={The {M}atlab {ODE} suite},
  author={Shampine, L. F. and Reichelt, M. W.},
  journal={SIAM J. Sci. Comput.},
  volume={18},
  number={1},
  pages={1--22},
  year={1997},
  publisher={SIAM}
}

@article{scotese2021phanerozoic,
  title={Phanerozoic paleotemperatures: The earth’s changing climate during the last 540 million years},
  author={Scotese, Christopher R and Song, Haijun and Mills, Benjamin JW and van der Meer, Douwe G},
  journal={Earth-Science Reviews},
  volume={215},
  pages={103503},
  year={2021},
  publisher={Elsevier}
}

@incollection{soderberg2024stratigraphy,
  author    = {Soderberg, Evan R. and Hart, Rachelle and Camp, Victor E. and Wolff, John A. and Steiner, Arron},
  title     = {Stratigraphy, Eruption, and Evolution of the Columbia River Basalt Group},
  booktitle = {Proterozoic Nuna to Pleistocene Megafloods: Sharing Geology of the Inland Northwest},
  editor    = {McFaddan, M. D. and Pritchard, C. J.},
  series    = {Geological Society of America Field Guide},
  volume    = {69},
  pages     = {81--121},
  year      = {2024},
  publisher = {Geological Society of America},
  doi       = {10.1130/2024.0069(05)}
}

@article{slyman2025tipping,
  title={Tipping mechanisms in a carbon cycle model},
  author={Slyman, K. and Fleurantin, E. and Jones, C. K. R. T.},
  journal={Chaos},
  volume={35},
  number={5},
  year={2025},
  publisher={AIP Publishing}
}

@article{walker2024oceanic,
  title={Oceanic anoxic event 2 triggered by kerguelen volcanism},
  author={Walker-Trivett, CA and Kender, S and Bogus, KA and Littler, K and Edvardsen, T and Leng, MJ and Lacey, J and Riding, JB and Millar, IL and Wagner, D},
  journal={Nat. Commun.},
  volume={15},
  number={1},
  pages={5124},
  year={2024},
  publisher={Nature Publishing Group UK London}
}

@article{wieczorek2011excitability,
  title={Excitability in ramped systems: the compost-bomb instability},
  author={Wieczorek, S. and Ashwin, P. and Luke, C. M. and Cox, P. M.},
  journal={Proc. R. Soc. A},
  volume={467},
  number={2129},
  pages={1243--1269},
  year={2011},
  publisher={The Royal Society Publishing}
}

@article{vasil2000evaluation,
  title={Evaluation of the volumes and genesis of Permian-Triassic trap magmatism on the Siberian Platform},
  author={Vasil’ev, Y. R. and Zolotukhin, V. V. and Feoktistov, G. D. and Prusskaya, S. N.},
  journal={Russian Geology and Geophysics},
  volume={41},
  number={12},
  pages={1645--1653},
  year={2000},
  publisher={Novosibirsk State University}
}

@article{wu2023volcanic,
  title={Volcanic {CO2} degassing postdates thermogenic carbon emission during the end-Permian mass extinction},
  author={Wu, Yuyang and Cui, Ying and Chu, Daoliang and Song, Haijun and Tong, Jinnan and Dal Corso, Jacopo and Ridgwell, Andy},
  journal={Sci. Adv.},
  volume={9},
  number={7},
  pages={eabq4082},
  year={2023},
  publisher={American Association for the Advancement of Science}
}

@article{zhang2018multiple,
  title={Multiple episodes of extensive marine anoxia linked to global warming and continental weathering following the latest {P}ermian mass extinction},
  author={Zhang, Feifei and Romaniello, Stephen J and Algeo, Thomas J and Lau, Kimberly V and Clapham, Matthew E and Richoz, Sylvain and Herrmann, Achim D and Smith, Harrison and Horacek, Micha and Anbar, Ariel D},
  journal={Sci. Adv.},
  volume={4},
  number={4},
  pages={e1602921},
  year={2018},
  publisher={American Association for the Advancement of Science}
}

@article{zeebe2003simple,
  title={A simple model for the CaCO3 saturation state of the ocean: The “Strangelove,” the “Neritan,” and the “Cretan” Ocean},
  author={Zeebe, Richard E and Westbroek, Peter},
  journal={Geochemistry, Geophysics, Geosystems},
  volume={4},
  number={12},
  year={2003},
  publisher={Wiley Online Library}
}

@article{BRYAN2008175,
title = {Revised definition of Large Igneous Provinces ({LIPs)}},
journal = {Earth-Sci. Rev.},
volume = {86},
number = {1},
pages = {175-202},
year = {2008},
issn = {0012-8252},
author = {Scott E. Bryan and Richard E. Ernst}}

@article{AlgeoShen2024,
  author  = {Algeo, Thomas J. and Shen, Jun},
  title   = {Theory and classification of mass extinction causation},
  journal = {National Science Review},
  year    = {2024},
  volume  = {11},
  number  = {1},
  pages   = {nwad237}
}

@article{BlackKarlstromMather2021,
  author  = {Black, Benjamin A. and Karlstrom, Leif and Mather, Tamsin A.},
  title   = {The life cycle of large igneous provinces},
  journal = {Nat. Rev. Earth Environ.},
  year    = {2021},
  volume  = {2},
  number  = {12},
  pages   = {840--857}
}

@article{BondGrasby2017,
  author  = {Bond, David P. G. and Grasby, Stephen E.},
  title   = {On the causes of mass extinctions},
  journal = {Palaeogeography, Palaeoclimatology, Palaeoecology},
  year    = {2017},
  volume  = {478},
  pages   = {3--29}
}

@article{burgess2015high,
  title={High-precision geochronology confirms voluminous magmatism before, during, and after Earth’s most severe extinction},
  author={Burgess, S. D. and Bowring, S. A.},
  journal={Sci. Adv.},
  volume={1},
  number={7},
  pages={e1500470},
  year={2015},
  publisher={American Association for the Advancement of Science}
}

@article{BurgessMuirheadBowring2017,
  author  = {Burgess, Seth D. and Muirhead, James D. and Bowring, Sam A.},
  title   = {Initial pulse of Siberian Traps sills as the trigger of the end-Permian mass extinction},
  journal = {Nat. Commun.},
  year    = {2017},
  volume  = {8},
  number  = {1},
  pages   = {164}
}

@article{BurgessBlack2025,
  author  = {Burgess, Seth D. and Black, Benjamin A.},
  title   = {The anatomy and lethality of the Siberian Traps large igneous province},
  journal = {Annu. Rev. Earth and Planet. Sci.},
  year    = {2025},
  volume  = {53},
  number  = {1},
  pages   = {567--594}
}

@article{DalCorsoEtAl2022,
  author  = {Dal Corso, Jacopo and Song, Haijun and Callegaro, Sara and Chu, Daoliang and Sun, Yadong and Hilton, Jason and Grasby, Stephen E. and Joachimski, Michael M. and Wignall, Paul B.},
  title   = {Environmental crises at the Permian--Triassic mass extinction},
  journal = {Nat. Rev. Earth Environ.},
  year    = {2022},
  volume  = {3},
  number  = {3},
  pages   = {197--214}
}

@article{DeeganEtAl2023,
  author  = {Deegan, Frances M. and Callegaro, Sara and Davies, Joshua H. F. L. and Svensen, Henrik H.},
  title   = {Driving global change one LIP at a time},
  journal = {Elements},
  year    = {2023},
  volume  = {19},
  number  = {5},
  pages   = {269--275}
}

@article{ErnstYoubi2017,
  author  = {Ernst, Richard E. and Youbi, Nasrrddine},
  title   = {How Large Igneous Provinces affect global climate, sometimes cause mass extinctions, and represent natural markers in the geological record},
  journal = {Palaeogeography, Palaeoclimatology, Palaeoecology},
  year    = {2017},
  volume  = {478},
  pages   = {30--52}
}

@article{GrasbyBond2023,
  author  = {Grasby, Stephen E. and Bond, David P. G.},
  title   = {How large igneous provinces have killed most life on Earth---Numerous times},
  journal = {Elements},
  year    = {2023},
  volume  = {19},
  number  = {5},
  pages   = {276--281}
}

@article{JoachimskiEtAl2020,
  author  = {Joachimski, M. M. and Alekseev, A. S. and Grigoryan, A. and Gatovsky, Y. A.},
  title   = {Siberian Trap volcanism, global warming and the Permian-Triassic mass extinction: New insights from Armenian Permian-Triassic sections},
  journal = {Geological Society of America Bulletin},
  year    = {2020},
  volume  = {132},
  number  = {1--2},
  pages   = {427--443}
}

@article{JonesEtAl2017,
  author  = {Jones, David S. and Martini, Anna M. and Fike, David A. and Kaiho, Kunio},
  title   = {A volcanic trigger for the Late Ordovician mass extinction? Mercury data from south China and Laurentia},
  journal = {Geology},
  year    = {2017},
  volume  = {45},
  number  = {7},
  pages   = {631--634}
}

@article{KaihoEtAl2021,
  author  = {Kaiho, Kunio and Aftabuzzaman, Md and Jones, David S. and Tian, Li},
  title   = {Pulsed volcanic combustion events coincident with the end-Permian terrestrial disturbance and the following global crisis},
  journal = {Geology},
  year    = {2021},
  volume  = {49},
  number  = {3},
  pages   = {289--293}
}

@article{KentEtAl2024,
  author  = {Kent, Dennis V. and Olsen, Paul E. and Wang, Huapei and Schaller, Morgan F. and Et-Touhami, Mohammed},
  title   = {Correlation of sub-centennial-scale pulses of initial Central Atlantic Magmatic Province lavas and the end-Triassic extinctions},
  journal = {Proceedings of the National Academy of Sciences},
  year    = {2024},
  volume  = {121},
  number  = {46},
  pages   = {e2415486121}
}

@article{LiEtAl2026,
  author  = {Li, Jing and Song, Huyue and Bond, David P. G. and Wignall, Paul B. and Du, Yong and Song, Haijun},
  title   = {On the causes of the end-Triassic mass extinction},
  journal = {Earth-Sci. Rev.},
  year    = {2026},
  pages   = {105542}
}

@incollection{MarzoliEtAl2017,
  author    = {Marzoli, A. and Callegaro, S. and Dal Corso, J. and Davies, J. H. and Chiaradia, M. and Youbi, N> and Bertrand, H. and Reisberg, L. and Merle, R. and Jourdan, F.},
  title     = {The Central Atlantic Magmatic Province {(CAMP)}: A review},
  booktitle = {The Late Triassic World: Earth in a Time of Transition},
  year      = {2017},
  pages      = {91--125},
  publisher={Springer International Publishing},
  editor={Tanner, L. H.}
}

@article{QiuEtAl2025,
  author  = {Qiu, Zhen and Kong, W. L. and Zhang, J. Q. and Kemp, David B. and Zhang, Qin and Liu, Wen and Grasby, Stephen E. and Zou, Caineng},
  title   = {Mercury evidences link intensive volcanism to the Late Ordovician mass extinction and changes in the atmosphere-land-ocean system},
  journal = {The Innovation Geoscience},
  year    = {2025},
  volume  = {3},
  number  = {2},
  pages   = {100124}
}

@article{SchoepferEtAl2022,
  author  = {Schoepfer, Stephen D. and Algeo, Thomas J. and Van de Schootbrugge, Bas and Whiteside, Jessica H.},
  title   = {The Triassic--Jurassic transition--A review of environmental change at the dawn of modern life},
  journal = {Earth-Sci. Rev.},
  year    = {2022},
  volume  = {232},
  pages   = {104099}
}

@article{VandeSchootbruggeWignall2016,
  author  = {Van de Schootbrugge, Bas A. S. and Wignall, Paul B.},
  title   = {A tale of two extinctions: converging end-Permian and end-Triassic scenarios},
  journal = {Geological Magazine},
  year    = {2016},
  volume  = {153},
  number  = {2},
  pages   = {332--354}
}

@article{Wignall2001,
  author  = {Wignall, Paul B.},
  title   = {Large igneous provinces and mass extinctions},
  journal = {Earth-Sci. Rev.},
  year    = {2001},
  volume  = {53},
  number  = {1--2},
  pages   = {1--33}
}

@article{WignallBond2024,
  author  = {Wignall, Paul B. and Bond, David P. G.},
  title   = {The great catastrophe: causes of the Permo-Triassic marine mass extinction},
  journal = {National Science Review},
  year    = {2024},
  volume  = {11},
  number  = {1},
  pages   = {nwad273}
}

@article{YangEtAl2025,
  author  = {Yang, Xiangrong and Yan, Detian and Wilson, David J. and Pogge von Strandmann, Philip A. E. and Liu, Xianyi and Liu, Chunyao and Tian, Hui and others},
  title   = {Lithium isotope and mercury evidence for enhanced continental weathering and intense volcanism during the Ordovician-Silurian transition},
  journal = {Geochimica et Cosmochimica Acta},
  year    = {2025},
  volume  = {391},
  pages   = {49--68}
}

@article{sudtip,
    author = {Sudakov, Ivan and Vakulenko, Sergey},
    title = {Bifurcations of the climate system and greenhouse gas emissions},
    journal = {Phil. Trans. R. Soc. A},
    volume = {371},
    number = {1991},
    pages = {20110473},
    year = {2013},
    month = {05},
   issn = {1364-503X},
    doi = {10.1098/rsta.2011.0473}}

@article{SUDAKOW202222,
title = {Knowledge gaps and missing links in understanding mass extinctions: Can mathematical modeling help?},
journal = {Phys. Life Rev.},
volume = {41},
pages = {22-57},
year = {2022},
issn = {1571-0645},
doi = {https://doi.org/10.1016/j.plrev.2022.04.001},
author = {Ivan Sudakow and Corinne Myers and Sergei Petrovskii and Colin D. Sumrall and James Witts}
}

@article{archer2009,
  author  = {Archer, David and Eby, Michael and Brovkin, Victor and Ridgwell, Andy and Cao, Long and Mikolajewicz, Uwe and Caldeira, Ken and Matsumoto, Katsumi and Munhoven, Guy and Montenegro, Alvaro and Tokos, Kathy},
  title   = {Atmospheric lifetime of fossil fuel carbon dioxide},
  journal = {Annu. Rev. Earth Planet. Sci.},
  volume  = {37},
  pages   = {117--134},
  year    = {2009},
  doi     = {10.1146/annurev.earth.031208.100206}
}

@article{cnsns2022excitable,
title = {Excitable media store and transfer complicated information via topological defect motion},
journal = {Commun. Nonlin. Sci.},
volume = {116},
pages = {106844},
year = {2023},
issn = {1007-5704},
doi = {https://doi.org/10.1016/j.cnsns.2022.106844},
author = {Ivan Sudakow and Sergey A. Vakulenko and Dima Grigoriev}
}

@article{
stewart2026,
author = {Emily M. Stewart  and Michael S. Diamond  and Lindsi J. Allman },
title = {Metamorphic sulfur release as a driver of sustained cooling and mass extinction},
journal = {Science Advances},
volume = {12},
number = {28},
pages = {eaee2277},
year = {2026},
doi = {10.1126/sciadv.aee2277}
}

\end{document}